\documentclass[preprint,times]{elsarticle}
\usepackage{hyperref}
\usepackage{amsmath}
\usepackage{caption}
\usepackage{subcaption}
\usepackage[margin=3.0cm]{geometry} % to fix Elsevier's roomy margins
\usepackage{placeins}   % to get table in Appendix B not mixed up with App. A

\usepackage[usenames,dvipsnames]{color}

\journal{Journal of Computational Physics}
\biboptions{numbers,sort&compress}

\def\xbar{\mathbf{\bar{x}}}
\def\xbf{\mbox{$\mathbf{x}$}}

\def\abf{\mbox{$\mathbf{a}$}}
\def\bbf{\mbox{$\mathbf{b}$}}

\def\Jbf{\mbox{$\mathbf{J}$}}

\DeclareMathOperator{\diag}{diag}
\numberwithin{equation}{section}
\DeclareMathOperator{\sign}{sign}

\begin{document}

\begin{frontmatter}

\title{SpECTRE: A Task-based Discontinuous Galerkin Code for Relativistic Astrophysics}

\author[cornell]{Lawrence~E.~Kidder}

\author[cornell,umass]{Scott~E.~Field}

\author[lbnl]{Francois~Foucart}

\author[pi,guelph,cct]{Erik~Schnetter}

\author[cornell]{Saul~A.~Teukolsky}

\author[cornell]{Andy~Bohn}

\author[cornell]{Nils~Deppe}

\author[cct,lsu]{Peter~Diener}

\author[cornell]{Fran\c{c}ois~H\'{e}bert}

\author[caltech]{Jonas~Lippuner}

\author[guelph,pi]{Jonah~Miller}

\author[caltech]{Christian~D.~Ott}

\author[caltech]{Mark~A.~Scheel}

\author[cita,toronto]{Trevor~Vincent}

\address[cornell]{Cornell Center for Astrophysics and Planetary
  Science, Cornell University, Ithaca, New York 14853, USA}

\address[umass]{Mathematics Department, University of 
  Massachusetts Dartmouth, Dartmouth, MA 02747, USA}

\address[lbnl]{Lawrence Berkeley National Laboratory, 1 Cyclotron Rd,
  Berkeley, CA 94720, USA; NASA Einstein Fellow}

\address[pi]{Perimeter Institute for Theoretical Physics, Waterloo,
  ON N2L 2Y5, Canada}

\address[guelph]{Department of Physics, University of Guelph, Guelph,
  ON N1G 2W1, Canada}

\address[cct]{Center for Computation \& Technology, Louisiana State
  University, Baton Rouge, LA 70803, USA}

\address[lsu]{Department of Physics \& Astronomy, Louisiana State
  University, Baton Rouge, LA 70803, USA}

\address[caltech]{TAPIR, Walter Burke Institute for Theoretical
  Physics, Mailcode 350-17, California Institute of Technology,
  Pasadena, CA 91125, USA}

\address[cita]{Canadian Institute for Theoretical Astrophysics,
  University of Toronto, Toronto M5S 3H8, Canada}

\address[toronto]{Department of Physics, University of Toronto, 60
  St. George Street, Toronto, ON M5S 3H8, Canada}

\begin{abstract}
We introduce a new relativistic astrophysics code, SpECTRE, that
combines a discontinuous Galerkin method with a task-based parallelism
model.  SpECTRE's goal is to achieve more accurate solutions for
challenging relativistic astrophysics problems such as core-collapse
supernovae and binary neutron star mergers.  The robustness of the
discontinuous Galerkin method allows for the use of high-resolution
shock capturing methods in regions where (relativistic) shocks are
found, while exploiting high-order accuracy in smooth regions.  A
task-based parallelism model allows efficient use of the largest
supercomputers for problems with a heterogeneous workload over
disparate spatial and temporal scales.  We argue that the locality and
algorithmic structure of discontinuous Galerkin methods will exhibit
good scalability within a task-based parallelism framework. We
demonstrate the code on a wide variety of challenging benchmark
problems in (non)-relativistic (magneto)-hydrodynamics. We demonstrate
the code's scalability including its strong scaling on the NCSA Blue
Waters supercomputer up to the machine's full capacity of $22,380$
nodes using $671,400$ threads.
\end{abstract}


\end{frontmatter}

%%%%%%%%%%%%%%%%%%%%%%%%%%%%%%%%%%%%%%%%%%%%%%%%%%%%%%%%%%%%%%%%%%%%%%%%%%%%%%%
\section{Introduction}
%%%%%%%%%%%%%%%%%%%%%%%%%%%%%%%%%%%%%%%%%%%%%%%%%%%%%%%%%%%%%%%%%%%%%%%%%%%%%%%

Numerical simulation of astrophysical phenomena is a computationally
challenging task. The relevant equations are often coupled nonlinear
partial differential equations (PDEs) with complicated
microphysics. High fidelity simulations necessarily require extremely
large computational grids in all three spatial dimensions on which
non-uniform workloads must be efficiently parallelized. A simulation
may involve large spatial and temporal dynamic ranges, and develop
(magneto-)hydrodynamic shocks and turbulent flows.

Accurate numerical simulations of astrophysical systems such as
neutron star mergers~\cite{eichler:89,1992ApJ...395L..83N,moch:93} and
core-collapse supernovae~\cite{janka:12review,burrows:13a,ott:16}
are crucial for achieving the full scientific potential of current and
future experiments such as the Fermi Gamma-Ray Space
Telescope~\cite{fermiweb} and the Laser Interferometer
Gravitational-Wave Observatory~\cite{ligo}.  Yet for many of these
systems the computational errors are often too large (or not even
quantifiable) with current algorithmic and hardware limitations.  The
simulations also take too long, several weeks to many months of wall
time on present supercomputers, precluding explorations of the
theoretical parameter space.
 
Within the astrophysics communities employing grid-based methods, the
industry standard has been finite-volume or finite-difference methods
parallelized by distributing cells across processors and communicating
data with message passing interface (MPI). The evolution is
synchronized according to a global simulation time. A variety of
astrophysics codes (e.g.,
Refs.~\cite{Fryxell:2000zz,bryan2014enzo,ott:16,mignone2011pluto,almgren2010castro,Loffler:2011ay,SpECwebsite})
have been designed based on these fundamental building blocks.

These strategies work well when the computations are reasonably
homogeneous or when one seeks good parallelization to only a few
thousand cores.  As the number of MPI processes increases, so does the
cost of communication which, together with non-uniform workload
typical of astrophysics problems, limits the maximum number of useful
cores that codes can run on.  Efficient core utilization becomes
non-trivial, often requiring careful optimization by hand to achieve
good scalability~\cite{woodward2013scaling}.
Standard finite-volume and
finite-difference methods achieve higher order accuracy
with increasingly large (overlapping) stencil sizes, 
and may require additional effort to achieve scalability 
on massively parallel machines.

As one looks ahead to the arrival of exascale computing, it will
become increasingly important to focus on developing algorithms that
can take full advantage of these very large machines.

Discontinuous Galerkin (DG)
methods~\cite{Reed.W;Hill.T1973,Hesthaven2008,Cock01,cockburn1998runge,Cockburn.B1998,Cockburn.B;Karniadakis.G;Shu.C2000},
together with a task-based parallelization strategy, have the
potential to tackle many of these problems.  DG methods offer
high-order accuracy in smooth regions 
(although, for stability, increasing the scheme's 
order requires decreasing the timestep, 
which restricts the largest usable order in practice),
robustness for shocks and other
discontinuities, and grid flexibility including a formulation that
allows for comparatively straightforward $hp$-adaptivity and local
timestepping. DG methods can be combined with 
positivity preserving
strategies~\cite{zhang2010positivity,balsara2016subluminal,balsara2012self}
or ``atmosphere treatments"~\cite{Muhlberger2014} 
which seek to maintain non-negative values of the pressure and density
in challenging regions such as those containing high-speed astrophysical flow.
DG methods are also well suited for parallelization:
Their formulation in terms of local, non-overlapping elements requires
only nearest-neighbor communication regardless of the scheme's order
of convergence.

Despite extensive success in engineering and applied mathematics
communities over the past two decades, applications in
relativity~\cite{Field:2010mn,brown2012numerical,field2009discontinuous,zumbusch2009}
and astrophysics~\cite{Radice:2011qr,mocz:14,zanotti:14,endeve:15}
have typically been exploratory or confined to simple problems.
Within the past year, however, there have been significant advances
toward production codes for
non-relativistic~\cite{schaal2015astrophysical} and
relativistic~\cite{teukolsky2015,Bugner:2015gqa} hydrodynamics,
special relativistic magnetohydrodynamics~\cite{zanotti2015}, and the
Einstein equations~\cite{Miller:2016vik}. These codes use MPI to
implement a data parallelism strategy.

In this paper, we describe SpECTRE, a general purpose discontinuous
Galerkin solver for relativistic astrophysics.  A distinguishing
feature of SpECTRE is its task-based parallelism model. Instead of
dividing work between parallel processes based on cell ownership, the
algorithm is decomposed into a list of tasks and their
inter-dependencies.  Examples of tasks include, for example, computing
a derivative in an element, computing a numerical flux on a boundary
or taking a time step.  Tasks are assigned to processes/threads
dynamically during the computation, in such a way as to satisfy
dependencies and to minimize the number of idle cores.  When a core
becomes idle, it is given another task to complete. This framework is
very different from the more traditional synchronous, data parallelism
model used in other grid-based astrophysics codes.

The algorithm's scalability is achieved through (i) separation of the
tasks of communication and computation, so that they can be
overlapped, (ii) asynchronous, non-blocking communication so that
cores are not idle, and (iii) a runtime system to manage task queues,
distribute tasks to cores, and gather timing statistics to inform
load-balancing decisions. The power of task-based parallelism has
already shown impressive performance in other application domains, for
example
Refs.~\cite{2016arXiv160602738S,phillips2005scalable,OpenAtomWebsite,jetley2008massively,uintah,berzins2010uintah,martyna2012openatom}.
SpECTRE uses the Charm++
library~\cite{CharmWebsite,kale1993charm++,shu1991chare,kale2016parallel}
to implement this parallelism model.

This paper is organized as follows.  The hydrodynamic systems
currently solved by SpECTRE are summarized in \S\ref{sec:laws}, and
include the non-relativistic Euler equation and the relativistic
(magneto-)hydrodynamics systems in arbitrary gravitational fields.
Next, in \S\ref{sec:galerkin} we describe a nodal DG scheme and those
approximate Riemann solvers and high-resolution shock capturing
limiters that we implement within SpECTRE. DG schemes naturally map
into a task-based parallelism framework, and in \S\ref{sec:tasks} we
describe how the algorithm can be broken down into tasks and
subsequently parallelized using Charm++.  Next, we present a sampling
of results for standard performance (see \S\ref{sec:performance}) and
benchmark (see \S\ref{sec:tests}) tests.  Our scalability experiments
demonstrate the power of a task-based approach. A key result is
Fig.~\ref{fig:StrongScaling-BWs}, which shows excellent strong
scalability on the Blue Waters machine up to its full capacity of
$22,380$ nodes using $671,400$ threads.

%%%%%%%%%%%%%%%%%%%%%%%%%%%%%%%%%%%%%%%%%%%%%%%%%%%%%%%%%%%%%%%%%%%%%%%%%%%%%%%
\section{Conservation laws}
\label{sec:laws}
%%%%%%%%%%%%%%%%%%%%%%%%%%%%%%%%%%%%%%%%%%%%%%%%%%%%%%%%%%%%%%%%%%%%%%%%%%%%%%%

SpECTRE is designed to solve conservation laws written in flux
conservative form.  A familiar example we shall consider is the
non-relativistic (or Newtonian) Euler equation of hydrodynamics.
Other conservation laws, such as the relativistic
(magneto-)hydrodynamics system, are posed on general spacetimes
equipped with a metric $g_{\mu \nu}$.  These systems are most easily
discussed with tensor notation. The details of these systems are not
necessary to understand the rest of the paper, which avoids the use of
tensors.

\subsection{Preliminaries and notation}

The metric in a general spacetime can be written in the standard space
plus time form~(See, e.g., \cite{baumgarteShapiroBook} or
\cite{rezzolla2013relativistic}),
\begin{equation}
ds^2 = g_{\mu\nu}dx^\mu dx^\nu
=-\alpha^2 dt^2+\gamma_{ab}(dx^a+\beta^a dt)(dx^b +\beta^b dt) \,.
\label{eq:3+1}
\end{equation}
Here $\alpha$ is called the lapse function, $\beta^a$ the shift
vector, and $\gamma_{ab}$ is the spatial metric on $t=\text{constant}$
hypersurfaces, sometimes called time slices.  Here and throughout,
repeated indices are summed over following Einstein's summation
convention.  In $d$ spatial dimensions, Greek indices $\mu,\nu,\ldots$
range from 0 to $d$, Latin indices $a,b,\ldots$ will be purely
spatial, ranging from $1$ to $d$.  In a flat spacetime (no
relativistic gravitational field), we can set $\alpha=1$, $\beta^a=0$.
Furthermore, in Cartesian coordinates, for a flat spacetime the
spatial metric is simply the Euclidean metric $\gamma_{ab}=
\delta_{ab}$.  Throughout this paper, we work in ``code units" in
which the speed of light $c = 1$ for all the relativistic applications
while for gravity applications Newton's constant $G=1$ and the solar
mass $M_{\odot} = 1$.

A conservation law posed on a generic spacetime takes the form
\begin{align} \label{eq:cons3}
& \frac{1}{\sqrt{\gamma}}\partial_t(\sqrt{\gamma}U) +
\frac{1}{\sqrt{\gamma}}\partial_a(\sqrt{\gamma}F^a) = S \,,
\end{align}
where $\gamma$ is the determinant of $\gamma_{ab}$.
We are interested in numerically solving~\eqref{eq:cons3} as an 
initial-boundary value problem
over the domain $\Omega$ 
subject to appropriate initial and boundary data.
Additional geometric factors present in Eq.~\eqref{eq:3+1} are hidden
inside the definitions of the state vector, $U$,
flux vectors, $F^a$, and source vector $S$. 
We denote the length of these vectors by $n$.
We now consider three particular systems 
that we implement in SpECTRE.

\subsubsection{Newtonian Euler hydrodynamics}

In $d$ spatial dimensions, the Euler equations form a set of $d+2$
coupled nonlinear conservation laws. These can be written in the more
general form~\eqref{eq:cons3} (with $\gamma = 1$ in Cartesian
coordinates) with an evolved state vector of conserved quantities
\begin{equation}
U  = \left[ \rho, \rho v_i, E \right]^T \,,
\end{equation}
flux vectors 
\begin{equation}
F^a = \left[ \rho v^a, \rho v_i v^a + p \delta^a_i, (E+p) v^a \right]^T \,,
\end{equation}
and (for the examples explored in this paper) a vanishing source vector $S$.
Here $\rho$ is the density, $E$ is the energy density, $p$ is the pressure, 
and $v^i = \gamma^{ij}v_j$($=v_i$ in Cartesian coordinates) are the components
of the fluid flow velocity. Let $\epsilon$ 
be the internal energy per unit mass. Then the energy can be written
as 
$E = \rho \epsilon + \frac{1}{2} \rho v^i v_i$.
The system is completed by an equation of state, which in general
can be written as $p=p(\rho, \epsilon)$ and may also depend on the 
fluid's composition. For the simple equations of state considered in this
paper, the relations between conserved and
primitive variables ($\rho$, $v_i$, $\epsilon$) are simple
algebraic expressions.

\subsubsection{Relativistic Euler hydrodynamics}
\label{sec:RelEuler}

In $d$ spatial dimensions, the relativistic Euler equations form a set
of $d+2$ coupled nonlinear conservation laws with the evolved state
vector of conserved quantities given by (e.g., \cite{lrr-2008-7})
\begin{equation}
\sqrt{\gamma} U =
\begin{pmatrix}
\tilde{D}\\ 
\tilde{S}_i\\
\tilde{\tau}
\end{pmatrix}
=
\begin{pmatrix}
\sqrt{\gamma} \rho W \\
\sqrt{\gamma} \rho h W^2 v_i \\
\sqrt{\gamma} \left( \rho h W^2 - p - \rho W\right) 
\end{pmatrix} \,,
\end{equation}
where the components of the fluid's $3$-velocity, 
$v^i = \gamma^{ij}v_j$, are defined from the 
fluid's $4$-velocity,  $u^\mu = W\left(1,v^i\right)$, 
$W\equiv \alpha u^0 = 1/\sqrt{1-v^i v_i}$ 
is the Lorentz factor, $h = 1+\epsilon+p/\rho$ is the 
relativistic specific enthalpy, and
$\epsilon$ is the specific internal energy in the fluid's rest frame. 
The pressure is given by a general equation of state $p(\rho, \epsilon)$.
Here a tilde denotes a ``densitized" version of the quantity, for example
$\tilde{D}=\sqrt{\gamma} D$.
Define the fluid ``transport velocity"
to be $v_{\rm tr}^i = \alpha v^i - \beta^i$.
Then the flux and source vectors are given by
\begin{equation}
\sqrt{\gamma} F^a =
\begin{pmatrix}
\tilde{D}v_{\rm tr}^a\\
\tilde{S}_iv_{\rm tr}^a+\sqrt{\gamma}\alpha p \delta^a_i\\
\tilde{\tau}v_{\rm tr}^a+\sqrt{\gamma}\alpha p v^a\\
\end{pmatrix}
,\qquad
\sqrt{\gamma} S =
\begin{pmatrix}
0\\
(\alpha/2) \tilde S^{lm} \partial_i \gamma_{lm} + \tilde S_k \partial_i \beta^k 
- \tilde E \partial_i \alpha\\
\alpha \tilde S^{lm}K_{lm} - \tilde S^l \partial_l \alpha\\
\end{pmatrix}
\,,
\end{equation}
with the source $\tilde S^{ij}$
and the energy $\tilde E$ given as
\begin{align}
\tilde S^{ij} &= \sqrt{\gamma} \rho h W^2 v^i v^j 
+ \sqrt{\gamma} p \gamma^{ij} \,,\\
\tilde E &= \sqrt{\gamma} \rho h W^2 - \sqrt{\gamma} p \,,
\end{align}
and where $K_{ab}$ is the extrinsic curvature tensor of the $t=$
constant slice and $K = \gamma^{ab} K_{ab}$ its trace.  For a flat
metric written in Cartesian coordinates ($\alpha=1$, $\beta^a=0$,
$\gamma_{ab}= \delta_{ab}$, $\gamma=1$, $K_{ab} = 0$, and $K = 0$) the
system simplifies to a set of conservation laws appropriate for
special relativistic hydrodynamics and the source vector $S$ is
identically zero.  We use the method outlined in Appendix C
of~\cite{Galeazzi:2013mia} to convert between conserved and primitive
variables ($\rho$, $v_i$, $\epsilon$).

\subsubsection{Relativistic magnetohydrodynamics}
\label{sec:rmhd}

The general relativistic equations of ideal magnetohydrodynamics
(GRMHD) can be written in conservative form similar to the equations
of hydrodynamics.  The main complication is the constraint in
Maxwell's equations that the divergence of the magnetic field
vanishes.  Constraint-satisfying initial data evolved according to
Maxwell's equations remain constraint satisfying when the evolution
equations are solved analytically. Small constraint violations can,
however, be numerically unstable. In order to avoid the growth of
unstable modes, we adopt an approach known as divergence
cleaning~\cite{Dedner2002} that is used in some finite-difference
codes~\cite{Palenzuela2009,Penner2011,moesta:14a}.  Divergence
cleaning introduces a new evolved variable, $\Phi$, that couples to
the divergence of the magnetic fields, and damps the unphysical mode
in the solution.

The exact form of the evolution equations that we use is a
modification of those presented in~\cite{moesta:14a}, in which we
recast the evolution equation for the divergence-cleaning scalar
$\Phi$ in conservative form.  Define $B^\mu$ as the magnetic field
measured by an observer moving along the normal to a $t=$ constant
slice (Eulerian observer) and $b^\mu$ as the magnetic field measured
by an observer comoving with the fluid. The two quantities are related
by the equations
\begin{align}
\alpha b^0 &= W B^i v_i \,, \\
b_i &= \frac{B_i}{W} + v_i \alpha b^0.
\end{align}
We also define a set of auxiliary variables
\begin{align}
b^2 &= b^\mu b_\mu \,, \\
(\rho h)^* &= \rho h + b^2 \,, \\
p^* &= p + \frac{b^2}{2}.
\end{align}
The evolved variables are then
\begin{equation}
\label{eq:evolved}
\sqrt{\gamma} U =
\begin{pmatrix}
\tilde{D}\\ 
\tilde{S}_i\\
\tilde{\tau}\\
\tilde{B^i}\\
\tilde{\Phi}
\end{pmatrix}
= \sqrt{\gamma}
\begin{pmatrix}
\rho W \\
(\rho h)^* W^2 v_i - \alpha b^0 b_i\\
(\rho h)^* W^2 - p^* - (\alpha b^0)^2-\rho W\\
B^i\\
\Phi
\end{pmatrix} \, .
\end{equation}
The flux and source vectors are given by
\begin{equation} \label{eq:MHD_RHS}
\sqrt{\gamma} F^j =
\begin{pmatrix}
\tilde D v_{tr}^j\\
\tilde S_i v_{tr}^j + \sqrt{\gamma}\alpha p^* \delta^j_i - \alpha \tilde B^j b_i/W\\
\tilde \tau v_{tr}^j + \alpha \sqrt{\gamma} p^* v^j - \alpha^2 b^0 \tilde B^j /W\\
v^j_{tr} \tilde B^i - \alpha v^i \tilde B^j + \alpha \gamma^{ij} \tilde \Phi\\
\alpha \tilde B^j - \tilde \Phi \beta^j
\end{pmatrix}
,\qquad
\sqrt{\gamma} S =
\begin{pmatrix}
0\\ 
(\alpha/2) \tilde S^{lm} \partial_i \gamma_{lm} + \tilde S_k
\partial_i \beta^k - \tilde E \partial_i \alpha\\ 
\alpha \tilde S^{lm}K_{lm} - \tilde S^l \partial_l \alpha\\ 
-\tilde B^j \partial_j \beta^i 
+ \Phi \partial_j (\alpha \sqrt{\gamma} \gamma^{ij}) \\ 
\alpha \tilde B^k \partial_k \ln{\alpha} - \alpha K \tilde \Phi - \alpha
\kappa \tilde \Phi
\end{pmatrix}
,
\end{equation}
with the source $\tilde S^{ij}$ and the energy density $\tilde E$
given as
\begin{align}
\tilde S^{ij} & = \sqrt{\gamma}\bigg((\rho h)^*W^2 v^i v^j + p^* \gamma^{ij}
- \gamma^{ik} \gamma^{jl} b_k b_l\bigg) \,, \\
\tilde E & = \tilde \tau + \tilde D \,.
\end{align}
We solve for the primitive variables using the algorithm proposed by
Newman and Hamlin~\cite{newman2014primitive}. Note that the exact
divergence-free (no-monopole) condition, $\Phi=\partial_i \tilde
B^i=0$, is analytically preserved by these equations, while
numerically the constraint violating mode will now be damped at a rate
$\kappa$.  The coefficient $\kappa$ is a damping parameter, which we
typically set in the range $\kappa \in [0.1,1]$.  Constraint violating
modes propagate at the speed of light.

\subsubsection{Treatment of unphysical conservative variables}
\label{sec:fixcons}

For the relativistic (magneto-)hydrodynamics systems, numerical error
can cause the conserved variables to evolve to values that do not correspond to any set of physical primitive variables.
In particular, this is likely to occur in low-density, highly relativistic, and/or strongly magnetized regions which do not
significantly affect the dynamics of the system when evolved exactly, but can cause significant numerical issues
if unphysical conserved variables are not handled properly. To avoid this issue, relativistic hydrodynamics codes generally
include {\it atmosphere corrections}, imposing a minimum density, maximum velocity, and maximum magnetization of the fluid
to stop low-density points from evolving towards unphysical values of the conserved variables. It is also common to include 
{\it conserved variable fixing}, i.e some prescription to handle rare points for which the conserved variables are unphysical.

The best prescriptions tend to be problem-dependent, and their discussion goes beyond the scope of this article. For the test
problems presented in this work, the atmosphere corrections only impose a positive density and internal energy. Fixes to
the conserved variables follow the methods outlined in Appendix A.2.c of~\cite{Muhlberger2014}. In that work, 
a maximum physical value of the momentum $S^iS_i$ and internal energy $\tau$ is defined. 
Here, the evolved variables are not allowed to grow closer than
one part in $10^{12}$ of those maximum values.
These simple atmosphere predictions would likely be insufficient when applying the code to large scale astrophysical simulations
(e.g. neutron star mergers, accretion disks), but work well for all test problems considered here.

%%%%%%%%%%%%%%%%%%%%%%%%%%%%%%%%%%%%%%%%%%%%%%%%%%%%%%%%%%%%%%%%%%%%%%%%%%%%%%%
\section{A nodal discontinuous Galerkin method}
\label{sec:galerkin}
%%%%%%%%%%%%%%%%%%%%%%%%%%%%%%%%%%%%%%%%%%%%%%%%%%%%%%%%%%%%%%%%%%%%%%%%%%%%%%%

\subsection{The algorithm}
\label{sec:algorithm}

\begin{figure*}
    \centering
    \def\svgwidth{1.0 \columnwidth}
    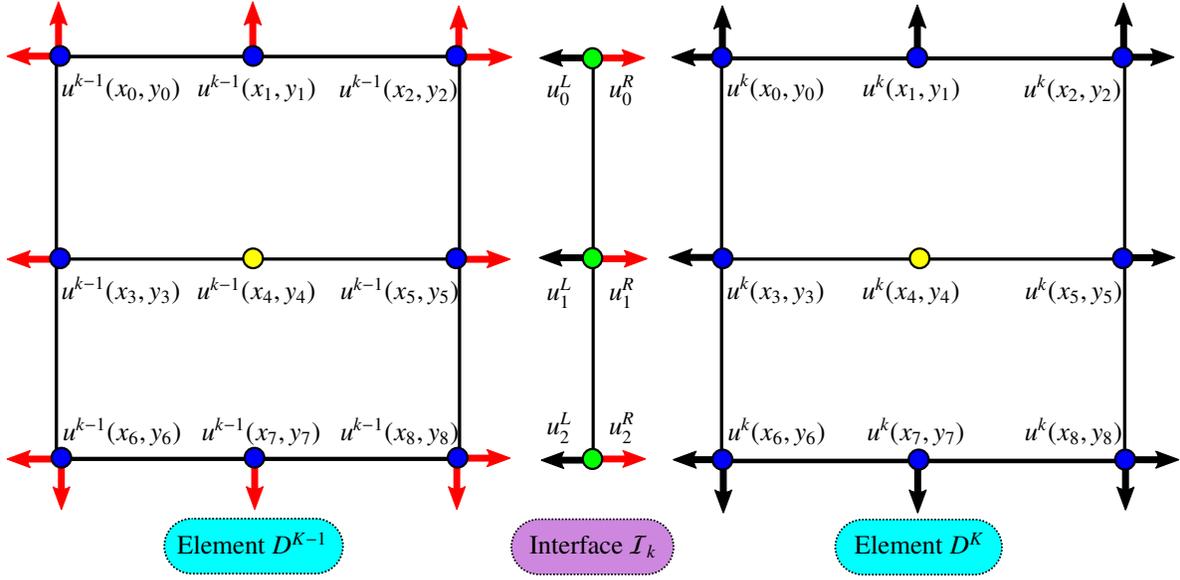
\caption{In two spatial dimensions the DG computational domain
  consists of a collection of elements (rectangles) with touching
  surfaces (lines) which we call interfaces. On each element the local
  numerical expansion gives function values at the set of GLL nodal
  points, shown here with second order elements.  While nodal points
  along each interface can be identical, for example the upper-right
  point in $D^{k-1}$ is the same as the upper-left point in $D^k$, the
  grid values may not be, $u^L_0 = u^{k-1}(x_2,y_2) \neq
  u^{k}(x_0,y_0) = u^R_0$.  Numerical fluxes are computed on each
  interface using grid data from two neighboring elements (blue
  circles) and local outward-pointing normals (red and black
  arrows). Our implementation of the DG scheme treats each interface
  and element as a C++ object equipped with tasks (see
  Fig.~\ref{fig:SpectreDecomposition}).
\label{fig:DGScheme}
}
\end{figure*}

Following Refs.~\cite{teukolsky2015,Hesthaven2008}, this section describes 
the nodal DG method we have implemented. The algorithm is derived using
the following steps (details in Ref.~\cite{teukolsky2015}):

\begin{itemize}
\item
Divide the spatial domain into elements.  Each element is a mapping of
a reference cube (in 3D) or square (in 2D) with extents $[-1,1]$ in
each direction.  The mapping is some time-independent function
\begin{equation}
\xbf=\xbf(\xbar)
\label{eq:12}
\end{equation}
with Jacobian matrix
\begin{equation}
\Jbf=\left(
\dfrac{\partial x^a}{\partial x^{\bar a}}
\right)
\label{eq:13}
\end{equation}
and Jacobian
\begin{equation}
J=\det\Jbf.
\label{eq:14}
\end{equation}
Here the barred coordinates are standard Cartesian-like coordinates
covering the reference element.
\item
In each element, each component of the quantities $\sqrt{\gamma}U$,
$\sqrt{\gamma} F^a$, and $\sqrt{\gamma} S$ is expanded in polynomial
basis functions.  We choose these basis functions to be a tensor
product of 1D basis functions $\ell_i$ on the reference element, so
that the expansion of a typical variable takes the form
\begin{equation}
\label{eqn:dgNodalInterpolation}
U(\xbar)=\sum_{ijk}U_{ijk}\ell_i\big(x^{\bar 1}\big)
  \ell_j\big(x^{\bar 2}\big)\ell_k\big(x^{\bar 3}\big)\,,
\end{equation}
where the time-dependent coefficients $U_{ijk}(t)$ are found from
Eq.~\ref{eq:411} below and $U_{ijk}(0)$ is given in terms of the
initial data (see \S\ref{sec:ID_BCS_steppers}).  The 1D basis
functions of degree $N$ are simply the Lagrange interpolating
polynomials corresponding to Legendre polynomials,
\begin{equation}
\ell_i(\bar x) = \prod_{\substack{j=0\\j\neq i}}^N
\frac{\bar x-\bar x_j}{\bar x_i-\bar x_j},
\end{equation}
where $\bar x_i$ are the nodes of a Gauss-Legendre-Lobatto (GLL) quadrature.
These nodes may be found with standard algorithms,
for example Algorithm 24 of Ref.~\cite{kopriva2009implementing}.

\item
In each element, follow the standard DG procedure of integrating
Eq.~\eqref{eq:cons3} multiplied by a basis function over the proper
volume $\sqrt{\gamma} d^3x$ of the element, where $d^3x$ is the
coordinate volume element $dx\,dy\,dz$,
\begin{equation}
\int\left[
\partial_t\left(\sqrt{\gamma} U\right) + \partial_a\left(
\sqrt{\gamma}F^a\right)-\sqrt{\gamma}S \right]
\phi_i(\xbf)\,d^3x=0.
\label{eq:3}
\end{equation}
Use integration by parts
(Gauss's Theorem) to convert
the divergence term to a surface integral:
\begin{align}
\int \partial_a\left(\sqrt{\gamma}F^a\right)\phi_i(\xbf)\,d^3x & =
\int \partial_a\left(\sqrt{\gamma}F^a\phi_i(\xbf)\right)\,d^3x
- \int \sqrt{\gamma}F^a\partial_a \phi_i(\xbf)\,d^3x\nonumber\\
&=\oint F^a n_a\phi_i\,d^2\Sigma
- \int \sqrt{\gamma}F^a\partial_a \phi_i(\xbf)\,d^3x.
\label{eq:3a}
\end{align}
Here $d^2\Sigma$ is the proper surface element of the cell and $n_a$ is
the unit outward normal.

With a formulation like (\ref{eq:3a}) in each element, there is no
connection between the elements.  The heart of the DG method is to
replace $F^a$ in the surface term by the numerical flux $F^{a*}$, a
function of the state vector on \emph{both} sides of the interface:
\begin{equation}
\int \partial_a\left(\sqrt{\gamma}F^a\right)\phi_i(\xbf)\,d^3x \to
\oint F^{a*} n_a\phi_i\,d^2\Sigma
- \int \sqrt{\gamma}F^a\partial_a \phi_i(\xbf)\,d^3x.
\label{eq:4}
\end{equation}
Now undo the integration by parts on
the right-hand side of Eq.\ (\ref{eq:4}):
\begin{equation}
\int \partial_a\left(\sqrt{\gamma}F^a\right)\phi_i(\xbf)\,d^3x
\to
\oint (F^{a*}-F^a) n_a\phi_i\,d^2\Sigma
+\int \partial_a\left(\sqrt{\gamma}F^a\right)\phi_i(\xbf)\,d^3x.
\label{eq:5}
\end{equation}

\item
Evaluate the integrals by using the expansion in basis functions,
mapping to the reference element with Eq.~\ref{eq:12}, and GLL
quadrature. The final result is Eq.~(3.17) of
Ref.~\cite{teukolsky2015}:
\begin{align}
\frac{d(\sqrt{\gamma}U)_{ijk}}{dt}  &+
\Big[
\frac{\partial x^{\bar 1}}{\partial x^a}\Big|_{ijk}\sum_l D_{il}^{\bar 1}
\left(\sqrt{\gamma}F^a\right)_{ljk}
+
\frac{\partial x^{\bar 2}}{\partial x^a}\Big|_{ijk}\sum_m D_{jm}^{\bar 2}
\left(\sqrt{\gamma}F^a\right)_{imk}\nonumber\\
&\quad+
\frac{\partial x^{\bar 3}}{\partial x^a}\Big|_{ijk}\sum_n D_{kn}^{\bar 3}
\left(\sqrt{\gamma}F^a\right)_{ijn}
\Big]-(\sqrt{\gamma}S)_{ijk}\nonumber\\
&=-
\frac{1}{w_N}F_{ijN}\frac{\sqrt{{}^{(2)}\gamma_{ij}}}{J_{ijN}}
 \delta_{kN}
+
\frac{1}{w_0}F_{ij0}\frac{\sqrt{{}^{(2)}\gamma_{ij}}}{J_{ij0}}
 \delta_{k0}
-
\frac{1}{w_N}F_{Njk}\frac{\sqrt{{}^{(2)}\gamma_{jk}}}{J_{Njk}}
 \delta_{iN}\nonumber\\
&\quad+
\frac{1}{w_0}F_{0jk}\frac{\sqrt{{}^{(2)}\gamma_{jk}}}{J_{0jk}}
 \delta_{i0}
-
\frac{1}{w_N}F_{iNk}\frac{\sqrt{{}^{(2)}\gamma_{ik}}}{J_{iNk}}
 \delta_{jN}
+
\frac{1}{w_0}F_{i0k}\frac{\sqrt{{}^{(2)}\gamma_{ik}}}{J_{i0k}}
 \delta_{j0}.
\label{eq:411}
\end{align}
Here $D^{\bar 1}_{il}$ is the differentiation matrix
\begin{equation}
D_{il}^{\bar 1}=\partial_{\bar 1} \ell_l\big(x^{\bar 1}\big)\big|_i
\label{eq:405}
\end{equation}
for $x^{\bar 1}$, and similarly for the 2- and 3-coordinates.
The quantity $F$ is the normal component of the flux difference,
$F=(F^{a*}-F^a) n_a$, where $F^{a*}$ is the numerical flux
and $n^a$ are the components of the element's locally outward-pointing 
normal vector (see Fig.~\ref{fig:DGScheme}).
The quantity ${}^{(2)}\gamma$ is the determinant of the
2-dimensional metric induced on the surface by $\gamma_{ij}$, and
$w_0$ and $w_N$ are the weights of the Gauss-Lobatto quadrature
at the endpoints of the interval.
\end{itemize}
We adopt a method--of--lines strategy and 
in Sec.~\ref{sec:ID_BCS_steppers} summarize those time steppers
we use to evolve the semi--discrete 
scheme~\eqref{eq:411}~\footnote{There is an 
  important technical issue that arises
  when implementing Eq.~\eqref{eq:411} in a code like SpECTRE. In
  Eq.~(3.18) of Ref.~\cite{teukolsky2015} it was shown that the
  boundary flux terms on the right-hand side of Eq.~\eqref{eq:411} can
  be simplified by using the \emph{unnormalized} normal vector when
  computing the fluxes.  However, we do not use this simplification
  and instead use the \emph{unit} normal $n_a$ explicitly.  The reason
  is that when the metric or mapping terms differ on the two sides of
  the boundary, as they will in curved spacetimes or with grid
  refinement, it is the unit normal that is the same on the two sides
  of the boundary, whereas pieces of the normalization factor like
  $\sqrt{{}^{(2)}\gamma}$ and $J$ differ.}.
 
Note that in the derivation of Eq.~\eqref{eq:411},  each product of
expansions is
evaluated using a single expansion with coefficients equal to the product
of the original coefficients. This replacement leads to an aliasing error:
contributions from the high order polynomials are aliased back onto
the basis.
While this does not affect the precision of the scheme,
it can lead to an aliasing-driven instability, which may
need to be dealt with by filtering~\cite{Hesthaven2008}.

In this paper, we will consider only the affine mapping
from the reference cube for each element:
\begin{equation} \label{eq:AffineMap}
\xbf(\xbar) = \abf^k + \tfrac{1}{2}(1+\xbar)(\bbf^k - \abf^k),
\end{equation}
where $\abf^k$ and $\bbf^k$ are the lower-left and upper-right coordinates
of the element. However, the code is designed to handle arbitrary
mappings.

\subsection{Numerical fluxes}

To complete the DG scheme we must specify functional forms for the
components of the numerical flux introduced in the previous subsection.
The numerical flux is determined by functions 
$F^{a*}(U^L,U^R)$ where
$U^L$ and $U^R$ are respectively the 
left and right (or lower and upper) boundary values
of the numerical approximation defined on 
neighboring elements (see Fig.~\ref{fig:DGScheme}).
To build a stable, convergent DG scheme the numerical 
flux must satisfy a few basic
properties~\cite{Hesthaven2008} such 
as consistency $F^{a*}(U,U) = F^{a}(U,U)$.
For smooth solutions, simple upwinding is a good prescription
for the numerical flux.
For non-smooth solutions, the flux is typically
prescribed by an approximate Riemann solver 
borrowed from the finite-volume community.
In both cases, the numerical flux
will, to varying degrees, make use 
of the characteristic decomposition of the 
conservation laws~\eqref{eq:cons3}.
In non-conservative 
form, this system of equations can be written as
\begin{equation}
\partial_t U + A^a(U) \, \partial_a U + \cdots = 0 \,,
\label{eq:NonConservative}
\end{equation}
where $A^a(U) = \partial_U F^a(U)$ is the flux Jacobian and the neglected
terms contain no derivatives of $U$. Since we assume that
the conservation law~\eqref{eq:cons3} gives a well-posed
initial-boundary value problem,
it follows that for any covector $n_a$ the matrix
\begin{equation}
n_a A^a(U) \equiv A(U;n) = K(U;n) \, \Lambda(U;n) \, K^{-1}(U;n)
\label{eq:CharDecomp}
\end{equation}
is diagonalizable. Here, 
$K(U;n)$ is an $n$-by-$n$ matrix whose $i^\mathrm{th}$ column
is the right eigenvector of $A(U;n)$ corresponding to 
the eigenvalue $\lambda_i$.
The eigenvalues (i.e. wavespeeds) are real, ordered
\begin{align*}
\lambda_1(U;n) \leq \lambda_2(U;n) \leq \ldots \leq \lambda_n(U;n) \,,
\end{align*}
and comprise the entries of the diagonal matrix $\Lambda(U;n) =
\diag\left( \lambda_1, \ldots, \lambda_n\right)$.  For the
(relativistic) hydrodynamics systems considered here, the
characteristic decompositions can be found in standard references
(e.g.~\cite{kulikovskii2000mathematical,alcubierreBook}).  For the
GRMHD system we are unaware of analytical expressions for the full
eigensystem, but the speeds can be computed as the roots of
polynomials given, e.g., in Ref.~\cite{Gammie:2003,DelZanna2007}.

If you apply the DG algorithm of \S\ref{sec:algorithm} to
Eq.~\eqref{eq:NonConservative}, you find that the analytic boundary
flux is
\begin{equation}
n_a F^a = n_a A^a U.
\end{equation}
The numerical flux is chosen to emulate this behavior,
with freedom to use varying amounts of information from the
characteristic decomposition.

We now summarize the numerical fluxes that we have implemented in our
code.  Later, numerical experiments will provide some insight into the
strengths and weaknesses of each choice.

\subsubsection{Local Lax-Friedrichs}

Because of its simplicity, robustness, and low
cost-of-evaluation, the local 
Lax-Friedrichs (LLF) numerical flux is a popular choice. 
However, it also introduces a large amount of dissipation
into the scheme for low polynomial orders
and does not use the system's wave structure,
which forms the building blocks of more sophisticated
approximate Riemann solvers. At each interface, 
the LLF flux is computed as
\begin{align}
\left(F^{a*} n_a \right)^{\rm LLF} =
\frac{F^{a}(U^L) n_a + F^{a}(U^R) n_a}{2} 
- \frac{\alpha^{\rm LF}}{2} \left(U^R - U^L \right) \,,
\label{eq:LLF}
\end{align}
where $\alpha^{\rm LF}$ is the maximum eigenvalue of either
$\left|\Lambda(U^R;n^R)\right|$ or
$\left|\Lambda(U^L;n^L)\right|$ computed at each 
collocation point along the interface. The LLF flux has a simple 
interpretation in terms of the average of the physical flux 
plus a dissipative part proportional to $\alpha^{\rm LF}$. Numerical dissipation
is necessary to stabilize the scheme, and setting $\alpha^{\rm LF} = 0$ 
is likely to be unstable, for example.

\subsubsection{Harten, Lax and van Leer}

The numerical flux proposed by
Harten, Lax, and van Leer (HLL)~\cite{HLL,toro2013riemann} is
\begin{align}
\left(F^{a*} n_a \right)^{\rm HLL} =
\frac{c_{\rm max}  F^{a}(U^L) n_a - c_{\rm min}  F^{a}(U^R) n_a }{c_{\rm max} - c_{\rm min}}
+ \frac{c_{\rm min} c_{\rm max} }{c_{\rm max} - c_{\rm min}} \left( U^R - U^L \right) \,,
\label{eq:HLL}
\end{align}
where $c_{\rm min}$ and $c_{\rm max}$ are estimates on the minimum
and maximum signal velocities bounding the left-moving and right-moving
wavespeeds that arise when solving the Riemann problem.
We use the simple, direct estimates 
proposed by Davis~\cite{davis1988simplified}
\begin{align}
c_{\rm min} = 
\min\left(\lambda_1\left( U^L\right), \lambda_1\left( U^R\right),0\right)
\,, \quad 
c_{\rm max} =
\max\left(\lambda_n\left( U^L\right), \lambda_n\left( U^R\right),0\right) \,.
\end{align}
Alternative estimates may perform better for some problems~\cite{toro2013riemann}.
Similar to the LLF flux,
the last term in Eq.~\eqref{eq:HLL} provides
a stabilizing diffusive term
(note that $c_{\rm min} \leq 0$ and $c_{\rm max} \geq 0$) while the 
first term is seen to be a weighted average. 
If $c_{\rm min} = 0$ or $c_{\rm max} = 0$, which will occur 
whenever all the characteristics are moving in the same direction,
the HLL flux reduces to pure upwinding. 

\subsubsection{Roe}

Roe's approach~\cite{Roe1981357,roe1985efficient} replaces the exact 
problem~\eqref{eq:NonConservative} by a linearized approximation
\begin{align}
\partial_t U + A_{\rm Roe}^a \, \partial_a U = 0 \,,
\label{eq:RoeApprox}
\end{align}
where the approximate flux Jacobian matrix, 
$A_{\rm Roe}^a(U^L,U^R)$, 
is required to satisfy conditions enforcing 
consistency, hyperbolicity, and conservation~\cite{toro2013riemann}.
If such a matrix $A_{\rm Roe}^a$ can be found, the 
Riemann problem~\eqref{eq:RoeApprox} can be exactly solved, 
and the numerical flux is given by
\begin{align}
\left(F^{a*} n_a \right)^{\rm Roe} =
\frac{F^{a}(U^L) n_a + F^{a}(U^R) n_a}{2} 
- \frac{1}{2} \sum_{i=1}^n \alpha_i \left| \lambda_i^{\rm Roe} \right| e_i^{\rm Roe} \,.
\label{eq:Roe}
\end{align}
Here $\lambda_i^{\rm Roe}$ and $e_i^{\rm Roe}$ are the $i^\mathrm{th}$
eigenvalue and right eigenvector of the Roe matrix, and the wave strengths 
\begin{align*}
\alpha_i = \left(E_i^{\rm Roe}\right)^T \left(U^R - U^L\right) \,,
\end{align*}
are computed from the projection of 
the state vector's jump onto the normalized left eigenvectors, $E_i$. 

For a non-relativistic ideal gas described by the Euler equations,  
a key result is that a Roe matrix can be constructed from
so-called Roe-averages,
\begin{equation}
\label{eq:RoeAverages}
\begin{gathered} 
\hat{u} = \frac{\sqrt{\rho_L}u_L + \sqrt{\rho_R}u_R}{\sqrt{\rho_L} + \sqrt{\rho_R}} \,, \quad
\hat{v} = \frac{\sqrt{\rho_L}v_L + \sqrt{\rho_R}v_R}{\sqrt{\rho_L} + \sqrt{\rho_R}} \,, \quad
\hat{w} = \frac{\sqrt{\rho_L}w_L + \sqrt{\rho_R}w_R}{\sqrt{\rho_L} + \sqrt{\rho_R}} \,, \\
\hat{h} = \frac{\sqrt{\rho_L}h_L + \sqrt{\rho_R}h_R}{\sqrt{\rho_L} + \sqrt{\rho_R}} \,, \quad
\hat{\rho} = \sqrt{\rho_L \rho_R}\,,
\end{gathered}
\end{equation}
through direct evaluation of the analytical flux Jacobian
at the Roe-averaged state
\begin{align} \label{eq:RoeMatrixIdealEOS}
A_{\rm Roe} = n_a A^a(\hat{u},\hat{v},\hat{w},\hat{h}) \, .
\end{align}
These expression have been written in the standard way
using $\left(u\,,v\,,w\right) \equiv \left(v^1\,,v^2\,,v^3\right)$ 
as the fluid flow velocity measured in a Cartesian coordinate system. 
It is easy to check that the matrix~\eqref{eq:RoeMatrixIdealEOS} satisfies
the three Roe conditions. For generic equations of state, a Roe matrix can
be constructed, but it will {\em not} be given by simple evaluation at 
a Roe-averaged
state~\cite{kulikovskii2000mathematical,glaister1988approximate}.
For the relativistic fluid equations, 
Eulderink and Mellema have constructed a 
Roe matrix that satisfies all
three Roe conditions~\cite{eulderink1994general}.
Nevertheless, we follow the accepted practice of defining a Roe-like 
matrix along the lines just described for 
a non-relativistic ideal gas. We expect such matrices to satisfy
the consistency and hyperbolicity conditions, but they are
unlikely to satisfy all of the Rankine-Hugoniot conditions.

\subsubsection{Marquina}

The flux proposed by Donat and Marquina~\cite{1996JCoPh.125...42D} also makes 
use of the characteristic decomposition of the problem 
but, as opposed to the Roe flux, does not require the computation of an intermediate state.
Instead, it uses the characteristic right eigenvectors $e_i^{R,L}$, normalized left
eigenvectors $E_i^{R,L}$ and eigenvalues $\lambda_i^{R,L}$ of the
linearized Jacobian matrices $A^a_{R,L}(U^{R,L})$ evaluated separately for the left and right
state vectors $U^{R,L}$. From these variables, we can define the projected variables and fluxes
\begin{align}
\omega^{R,L}_i = \left(E_i^{R,L}\right)^T U^{R,L} \,, \quad \phi_i^{R,L} = \left(E_i^{R,L}\right)^T F^a(U^{R,L})n_a.
\end{align}
The Marquina numerical flux is then defined as
\begin{align}
(F^{a*}n_a)^{\rm Marquina} = \sum_{i=1}^n \left(\phi_i^+ e^i_L + \phi_i^- e^i_R \right).
\end{align}
For the i$^{\rm th}$ eigenvector, the fluxes $\phi_i^\pm$ are defined so that the scheme is upwind
if $\lambda_i^R \lambda_i^L \geq 0$, but switches to a more dissipative LLF scheme
otherwise. That is, if $\lambda_i^{R,L} \geq 0$, we set $(\phi_i^+ = \phi_i^L\,;\phi_i^-=0)$.
If $\lambda_i^{R,L} \leq 0$, we choose $(\phi_i^- = \phi_i^R\,;\phi_i^+=0)$. Otherwise, the LLF
flux is obtained using
\begin{align}
\alpha_i = \max{(|\lambda_i^R|,|\lambda_i^L|)}\,, \quad \phi_i^+ = \frac{1}{2}(\phi_i^L+\alpha_i \omega^L_i) \,, \quad
\phi_i^- = \frac{1}{2}(\phi_i^R-\alpha_i \omega^R_i).
\end{align}
While the Marquina flux has the advantage of avoiding the intermediate state used by the Roe flux,
it does require the computation of the eigenvectors and eigenvalues of the characteristic matrix
for both the left and right state at each interface.
For the relativistic hydrodynamics system, 
expressions for both sets of eigenvectors
can be found, e.g., in Refs.~\cite{1999astro.ph.11034I,lrr-2008-7}.

\subsubsection{Numerical dissipation and maximum wave speeds for the relativistic MHD system}

As we have just seen, Riemann solvers which do not make use of the
characteristic eigenvectors of the problem (e.g., the LLF and HLL
numerical fluxes) introduce numerical dissipation proportional to the
maximum characteristic speed of the system.  For the relativistic MHD
system described in Sec.~\ref{sec:rmhd}, we do not have analytical
expressions for the characteristic eigenvectors and characteristic
speeds of the system. We do know, however, that the use of a
divergence cleaning method introduces the speed of light as one of the
characteristic speeds.

Using the speed of light as the maximum speed in the HLL/LLF fluxes
would introduce a large amount of dissipation, while using a smaller
characteristic speed in the dissipative component of those fluxes is
unstable. Yet, it is easy to see that the fluxes $F^l$ for the
divergence cleaning variable $\tilde \Phi$ and the longitudinal
component of the magnetic field $\tilde B^l$ are functions of $\tilde
B^l$ and $\tilde \Phi$, but not of the other evolved variables of the
GRMHD equations.  Accordingly, at each interface, we can separate the
GRMHD equations into two sub-systems: a system of 2 equations for the
divergence cleaning variable and the longitudinal component of the
field (component normal to the interface), and a system of 7 equations
for all other variables. The characteristic eigenvalues for the small
system $(\tilde B^l,\tilde \Phi)$ are $\lambda_\pm = -\beta^l \pm
\alpha \sqrt{\gamma^{ll}}$, and the corresponding eigenvectors are
$(\pm \sqrt{\gamma^{ll}},1)$. This sub-system can be treated
separately from the larger sub-system of 7 equations. For example,
we could use the HLL/LLF fluxes with the characteristic speeds
$\lambda_\pm$, or
the Roe or Marquina numerical fluxes (for
$\beta^l=0$ and $\alpha$ continuous across the interface, the Marquina
and HLL fluxes are in fact identical in this case).

The larger sub-system of 7 equations has its own complete set of 7
eigenvectors and eigenvalues, and its eigenvectors are orthogonal to
those of the smaller sub-system $(\tilde B^l,\tilde \Phi)$. The system
also has the same eigenvalues and eigenvectors as the GRMHD equations
evolved without divergence cleaning. While we do not know the maximum
eigenvalues of that system analytically, we can either solve for them
numerically or make use of the known bound on the value of the
characteristic speeds in the fluid frame~\cite{Gammie:2003}
\begin{equation}
\lambda_{\rm fl}^2 \leq \left( v_A^2  + c_s^2 (1-v_A^2)\right),
\end{equation}
with $c_s$ the relativistic sound speed and $v_A$ the {\em Alfven} speed.

In the tests presented here, we use the bound on $\lambda_{\rm fl}^2$
to determine an approximate minimum and maximum characteristic speed
in the grid frame. We use the same method for the computation of the
fluxes for both subsystems (either HLL of LLF), but with a different
choice of maximum characteristic speed for the sub-system $(\tilde
B^l,\tilde \Phi)$, thus recovering stability without adding
dissipation to the evolution of the other variables.
In the future we hope to explore  
recently developed Riemann solvers for the relativistic 
MHD system~\cite{punsly2016riemann,dumbser2016new} 
which may offer interesting advantages as compared  
to the more traditional numerical flux prescriptions 
we have considered here.

\subsection{High-resolution shock capturing limiters}
\label{sec:HRSCLimiters}

It is well known that an approximation of a non-smooth function
by smooth polynomials will produce spurious (Gibbs) oscillations 
with overshoots. In these troublesome elements the DG scheme's
order of convergence will be reduced and the solution may
differ qualitatively from the true solution. For example, if 
negative densities are generated from undershoots,
the wavespeeds can become imaginary, triggering an instability. 

The main challenge is to modify the numerical solution in such a way that
the spurious oscillations are removed (or at least acceptably controlled)
while retaining as much accuracy as possible. This is the job of 
{\em limiters}. As an extreme example,
the modification rule could be to set the solution equal to its
cell average, equivalent to retaining only the constant part of the 
polynomial basis. Such a heavy-handed limiter is guaranteed to remove
spurious oscillations but at the price of reducing the scheme's
accuracy to first order. 
More sophisticated limiting procedures must 
be able to discriminate between spurious oscillations
generated near non-smooth features
and genuine maxima or minima of the true solution.
For nonlinear hyperbolic partial differential 
equations, an additional complication arises as the solution
may develop physical discontinuities. A robust limiter should be 
able to handle a diverse range of problems without the need for
fine-tuned numerical parameters.

Given the importance of the problem, extensive work has gone into the
development of high-resolution shock capturing (HRSC) limiters.
As the name suggests, HRSC limiters seek to retain the accuracy afforded 
by the polynomial basis while resolving sharp features 
such as shocks. They typically involve two conceptually distinct parts.
First, a {\em troubled-cell indicator} is used to identify
elements that may contain spurious oscillations. Next, 
a {\em limiter} modifies the solution to reduce or 
completely remove the offending oscillation. 
When implemented within the Method of Lines
framework, this procedure is applied to both the advanced
solution and any intermediate stages required by the timestepper.

We now summarize those HRSC limiters that we have implemented.
These limiters can all be applied straightforwardly to either the set of
conserved, primitive or characteristic variables; in the literature one can 
find endorsements for all three 
(e.g.~\cite{schaal2015astrophysical,cockburn1998runge,Hesthaven2008,moe2015simple}).
Unless noted otherwise, our discussion and application of limiters will 
be applied to the conserved variables. We also note
that we have found it more stable to apply these limiters {\it before}
any correction is applied to the conserved variables (see Sec.~\ref{sec:fixcons}),
and to always flag cells in which the conserved variables are unphysical 
as troubled. In particular, the {\it FastShock} test described in Sec.~\ref{sec:testrmhd} 
does not converge to the correct solution unless unphysical conserved variables
are flagged as troubled and limited.

\subsubsection{Slope limiters}

Slope limiters were originally introduced in finite-volume methods and
were some of the earliest attempts at doing better than setting the
troubled cell to its average.  These limiters are generally {\em at
  best} second-order accurate and work by (i) modifying the solution's
slope and (ii) dropping any higher-order terms originally present in
the approximate solution~\eqref{eqn:dgNodalInterpolation}. We will
consider a family of minmod-based
limiters~\cite{cockburn1998runge,qiu2005comparison,Hesthaven2008} that
replace the numerical solution in element $D^k$ by a piecewise linear
representation:
\begin{equation} \label{eq:MinModSol}
U^k \rightarrow U^k_0
+ a_x \left(x - x_0\right)
+ a_y \left(y - y_0\right) 
+ a_z \left(z - z_0\right) \,.
\end{equation}
Here $a_x$, $a_y$, $a_z$ are estimates of the solution's slope along the 
three coordinate directions, $U^k_0$ is the solution's mean value,
and $\left(x_0, y_0, z_0\right)$ is the coordinate
center of the element.

We build Eq.~\eqref{eq:MinModSol} using data from $D^k$ and its six
neighbors by considering each dimension separately. Consider the task
of finding $a_x$.  From $D^k$ we use the numerical solution's slope,
$U^k_x$, and cell average, $U^k_0$. From its left and right neighbors,
$D^{k-1}$ and $D^{k+1}$, we require the numerical solution's cell
averages, $U^{k-1}_0$ and $U^{k+1}_0$. Next, we compute three
estimates of the slope in the $x$-direction:
\begin{equation}
a_1 = U^k_x \,, \quad a_2 = \frac{U^{k+1}_0 - U^k_0 }{\Delta x / \beta}
\,, \quad a_3 = \frac{ U^k_0 - U^{k-1}_0}{\Delta x / \beta} \,,
\end{equation}
where $\Delta x$ is the element's length and $\beta = 2$ 
for standard minmod~\cite{cockburn1999discontinuous,Hesthaven2008}
and $\beta = 1$ for the more 
dissipative MUSCL (Monotone Upstream Centered Scheme for Conservation Laws)
limiter of
van Leer~\cite{Hesthaven2008,van1974towards,van1979towards,cockburn1999discontinuous}.
If the solution were known 
to be smooth enough within $D^k$, we would
hope to retain $a_x = U^k_x$ as the slope
since it would be the most accurate. For smooth solutions
and not-too-large values of $\Delta x$, 
the values $a_2$ and $a_3$ are expected to be 
reasonable estimates of the true slope. 
Spurious oscillations (and, unfortunately, local maxima and minima)
manifest themselves as inconsistent signs of the three estimates.
These considerations motivate the introduction of a 
{\em minmodB} function
\begin{equation} \label{eq:MinModB}
a_x = m_B(a_1,a_2,a_3;M) \equiv 
\begin{cases}
a_1,
& \left| a_1 \right| \leq M \left(\Delta x\right)^2 \,,\\
m(a_1,a_2,a_3), & \text{otherwise}.
\end{cases}
\end{equation}
where the {\em minmod} function is
\begin{equation} \label{eq:MinMod}
m(a_1,a_2,a_3) = 
\begin{cases}
\sign\left(a_1\right) \min{\left(|a_1|, |a_2|, |a_3|\right)},
& \sign\left(a_1\right) = \sign\left(a_2\right) = \sign\left(a_3\right) \,, \\
0, & \text{otherwise}\,.
\end{cases}
\end{equation}
The minmodB function $m_B$ returns a value that we take to be 
the estimated slope $a_x$.
The positive constant $M$ controls the amount of oscillation we 
are willing to tolerate. 
If $M=0$, the slope limiting procedure guarantees that the numerical solution 
will be total variation diminishing in the mean (i.e., no spurious
oscillations in the cell averages), but necessarily 
clips off the solution's extrema. When $M \neq 0$ we allow local
maxima and minima to potentially escape limiting, but
continue to suppress large, spurious oscillations. Ideally
$M$ would be chosen proportional to the solution's 
second derivative at the extrema, although in practice 
its value may not require any special fine-tuning~\cite{qiu2005comparison}.
Estimated values for the slopes $a_y$ and $a_z$ follow the exact same recipe.

To decide whether the solution is in need of limiting, 
we (i) project the solution onto the space of piecewise linear functions
and then (ii) modify the linearized solution's slope using the minmodB
function. Notice that the minmodB slope limiter acts as both 
a troubled cell detector and limiter; for 
a piecewise linear basis no
limiting occurs whenever $a_1 = m_B(a_1,a_2,a_3;M)$. 
For a piecewise linear basis this is reasonable, whereas for higher-order
bases we use a generalization described next. 

\subsubsection{Generalized high-order minmodB limiter}

Whenever our basis functions are of degree $N \geq 2$, we use a
high-order generalization of the minmodB limiter
as a troubled cell detector. Considering one-dimension 
and following Ref.~\cite{Hesthaven2008}
(see also Refs.~\cite{cockburn1998runge,cockburn1999discontinuous}),
we first compute the values 
\begin{align*}
v_l & = U^k_0 - m_B(U^k_0 - U^k(x(-1)) , U^k_0 - U^{k-1}_0,U^{k+1}_0 - U^k_0;0) \,, \\
v_r & = U^k_0 + m_B(U^k(x(1)) - U^k_0 , U^k_0 - U^{k-1}_0,U^{k+1}_0 - U^k_0;0) \,.
\end{align*}
If the numerical solution satisfies both conditions 
$\left| v_l  - U^k(x(-1)) \right| \leq 10^{-8}$ and 
$\left| v_r  - U^k(x(1))  \right| \leq 10^{-8}$
then the full
high-degree polynomial solution is used and no limiting is performed.
Otherwise, the numerical solution is modified through slope limiting: 
we (i) linearize
\begin{equation}
U^{\rm Lin} = U^k_0 + U^k_x \left(x - x_0\right) \,,
\end{equation}
by dropping all of the higher-order terms 
originally present in the numerical solution~\eqref{eqn:dgNodalInterpolation},
(ii) apply the minmodB slope limiter (with $M \neq 0$) to $U^{\rm Lin}$ and finally
(iii) replace the numerical solution with the minmodB limited solution computed in step (ii).
Here, for example, $U^k(x(1))$ is the numerical solution 
evaluated at the reference element's coordinate value $\bar{x} =1$,
which labels the right-side boundary of $D^k$. 

To extend this procedure into higher dimensions we simply replace the two
function evaluations, $U^k(x(-1))$ and $U^k(x(1))$, with mean values of the 
solution computed over the element's face perpendicular to the outward-pointing
normal $\hat{x}$. Limiting is performed if 
the numerical solution fails to satisfy any one of the now six conditions 
tested in the $x$, $y$, and $z$ directions. 

\subsubsection{Limiter of Moe, Rossmanith and Seal}

Moment limiters extend the idea of modifying the
solution's slope information to their higher-moment analogues, like
those terms in the approximation~\eqref{eqn:dgNodalInterpolation} 
proportional to $x^3$, $xyz$, $yz^2$ etc. We will consider 
a recently proposed moment limiter of
Moe, Rossmanith and Seal (MRS)~\cite{moe2015simple} which has the following
desirable properties: in smooth regions it retains the full 
high-order accuracy afforded by the basis, its implementation requires
minimal communication between neighbors, and 
it has demonstrated good performance on a variety of 
benchmark tests~\cite{moe2015simple}.

Instead of the ansatz~\eqref{eq:MinModSol}, we now seek
to replace the numerical solution as follows:
\begin{equation} \label{eq:MRSSol}
U^k \rightarrow U^k_0 + \Theta_k \left(U^k - U^k_0\right) \,.
\end{equation}
This replacement
rescales all of the higher-order moments by the rescaling
function $\Theta_k \in \left[0,1\right]$. In the presence 
of large spurious oscillations we expect $\Theta_k=0$, and the 
numerical solution is replaced by its cell average. In smooth
regions we expect $\Theta_k=1$, and the solution remains
unaltered. To implement the MRS limiter, we follow the steps 
described in Ref.~\cite{moe2015simple} and, in particular, for
systems of equations we use the primitive variables to find $\Theta_k$ whereas
the rescaling~\eqref{eq:MRSSol} is carried out on the conserved
variables, thereby retaining conservation.

The effectiveness of this limiter depends on a free
parameter, $\tilde \alpha \geq 0$, which is used to estimate
the solution's upper and lower bound 
on each element. Following Ref.~\cite{moe2015simple}, 
we compute these estimates from neighbor data and 
the value of $U^k_0 \pm \tilde \alpha h^{3/2}$
(we take $\alpha(h) = \tilde \alpha h^{3/2}$
in Eqs.~3.1 and 3.2 of Ref.~\cite{moe2015simple}).
Here $h = \Delta x$ is the width of a uniform
element and the bounds are computed separately for each component of the 
vector of primitives. A non-zero value of $\tilde \alpha$ is required to retain
high-order accuracy near smooth extrema while $\tilde \alpha=0$ 
suppresses both extrema and spurious oscillations. In practice
we find a wide range of values are effective for a given problem. 
As shown in Ref.~\cite{moe2015simple}, for any non-zero value 
of $\tilde \alpha$ the MRS limiter will turn off around smooth
extrema as the grid is refined $h \rightarrow 0$.

\subsection{Initial data, boundary conditions and timesteppers}
\label{sec:ID_BCS_steppers}

Spatial approximation of the underlying system by the DG strategy
leads to Eq.~\eqref{eq:411}. Time integration of this equation
can then be carried out with a suitable ODE integrator, that is, we
are using the method-of-lines.
We have implemented two third-order integrators:
a multi-step Adams-Bashforth (AB3)
and a multi-stage strong stability-preserving 
Runge-Kutta timestepper~\cite{gottlieb1998total,gottlieb2001strong}. 
A CFL condition restricts the
largest stable timestep $\Delta t_{\rm max}$
associated with explicit numerical integration of
Eq.~\eqref{eq:411}. For a DG scheme, it is known that 
$\Delta t_{\rm max}$ is inversely proportional to the largest 
wavespeed, $\lambda_{\rm max}$, 
arising from the characteristic decomposition~\eqref{eq:CharDecomp} and
proportional to the smallest distance between
neighboring GLL points on the physical grid, $\Delta x_{\rm min}$. 
Since the value of $\Delta t_{\rm max}$ in fact 
depends on all aspects of the scheme, 
such as details of the numerical flux and the timestepper's stability region,
we find a stable $\Delta t$ from the scaling relation
\begin{align*}
\Delta t_{\rm max} \sim \mathrm{C} \frac{ \Delta x_{\rm min}}{\lambda_{\rm max}} \,,
\end{align*}
and a reasonable estimate for the unknown scaling factor $\mathrm{C}$.

Initial data for Eq.~\eqref{eq:411}
is provided by interpolating onto the numerical grid. 
At the interface located on the domain's physical outer boundary, we 
use the same numerical flux used for interior interfaces
but now supply the exterior solution. Throughout this paper we use 
periodic or analytic physical boundary conditions for the exterior state.

\subsection{Computational considerations}
\label{sec:comp_costs_main}

The time-to-solution is primarily determined by three different cost
factors: computation, data movement, and inter-process communication.

\emph{Computation} is probably the most well known of these.
Computational cost is often measured in Flop (floating-point
operations), and speed is measured in Flop/s (floating-point
operations per second). A high-performance compute node can today
execute about $1000\,\mathrm{GFlop/s}$. Of course, such a compute
speed can only be obtained if there are no other bottlenecks in the
code, and usually there are.

\emph{Data movement} is today for many algorithms the actual limiting
factor. A core's registers and caches can hold only a small amount of
data, and accessing the main memory is slow, both in the sense of
large latencies and small bandwidths, when compared with the
theoretical computational peak performance of a node.
A typical latency for loading data from memory is of
the order of $100\,\mathrm{ns}$, and a typical bandwidth is of the
order of $10\,\mathrm{GByte/s}$ to $100\,\mathrm{GByte/s}$. This
memory bandwidth is shared between all the cores of a socket.

The relevant property of a computational algorithm is its
\emph{byte-per-Flop} ratio, i.e., the average (amortized) number of
bytes that need to be loaded from or stored to memory for each
floating-point operation. Many algorithms have byte-per-flop ratios of
about $1\,\mathrm{Byte/Flop}$, whereas many computing systems have
byte-per-flop ratios of only about $0.1\,\mathrm{Byte/Flop}$. In other
words, performance (even under the idealized assumptions) is 
often limited by the available memory bandwidth, and not by the
available compute power.

Instead of letting the excess compute power go to waste, one can
use it to perform additional computations on data already loaded from
memory. For example, when limited by data movement costs, 
the additional accuracy of a high-order DG algorithm
almost comes for free.

\emph{Inter-process communication} arises from the need to communicate
between different processes running on different nodes.  Today, all
large HPC systems employ a distributed memory architecture requiring
inter-process communication. A modern HPC system (using, for example,
an InfiniBand network) has a bandwidth on the order of
$1\,\mathrm{GByte/s}$ to $10\,\mathrm{GByte/s}$ and latencies of
$1\,\mu s$ to $10\,\mu s$.  The Blue Waters' Gemini interconnect,
which we use in our scalability experiments, achieves a peak bandwidth
of $9.6\,\mathrm{GByte/s}$.

In a spatial domain decomposition with a homogeneous workload, the
amount of work per element scales with the volume of the element,
while the amount of data that needs to be communicated scales with its
surface area. The communication cost of an algorithm depends on the
``thickness'' of the surface layer that needs to be communicated. By
construction, the DG algorithm only requires communication of
interface data (a thickness of one) regardless of the scheme's order
of convergence (see Figure~\ref{fig:DGScheme}).  For example, suppose
a 3D domain has $N$ elements in total. When distributed over $P$
processes, the computation and communication cost scale as $O \left(
N/P \right)$ and $O \left( (N/P)^{2/3} \right)$, respectively.  The
communication-to-computation ratio is then $O \left( (P/N)^{1/3}
\right)$. This limits the number of processes over which a problem
with a given size can be distributed; once the communication cost is a
significant fraction of the overall cost, increasing $P$ will increase
the total cost.

Network latency also contributes to the communication cost. The
latency can be ignored if the communication time is much larger than
the latency. This is the case if the amount of communicated data is
larger than the network's \emph{latency-bandwidth-product}.  Using the
numbers quoted above, current HPC networks have a
latency-bandwidth-product of the order of $10\,\mathrm{kByte}$.  If
typical messages are shorter than this, then one way to hide the
latency is to employ a multi-threading system where multiple tasks can
initiate independent communications simultaneously, so that some tasks
can continue doing useful work while other tasks are blocked on
communication.

DG methods are particularly well-suited for parallelization. Their
formulation in terms of local, non-overlapping elements requires only
nearest-neighbor communication of surface data terms. They achieve
higher-order accuracy (which increases the compute cost per element)
without increasing the communication cost, unlike, e.g.,
finite-difference methods where the stencil size grows with the order
of accuracy. 

%%%%%%%%%%%%%%%%%%%%%%%%%%%%%%%%%%%%%%%%%%%%%%%%%%%%%%%%%%%%%%%%%%%%%%%%%%%%%%%
\section{Task-based parallelism}
\label{sec:tasks}
%%%%%%%%%%%%%%%%%%%%%%%%%%%%%%%%%%%%%%%%%%%%%%%%%%%%%%%%%%%%%%%%%%%%%%%%%%%%%%%

\subsection{General considerations}

The de facto standard for parallel and distributed programming to
solve large systems of PDEs is based on MPI and multi-threading,
usually implemented using OpenMP. The simulation domain is split onto
MPI processes and data are communicated across the element
boundaries. This approach has the advantage that it is rather
straightforward to implement, and it scales well if the computations
are reasonably homogeneous. The hallmarks of this approach are
globally synchronous phases, alternating between computation and
communication.

If different regions of the domain require different calculations,
then the resulting \emph{load imbalance} can be difficult to address. In
addition, the global communication phases lead to delays that are in
principle avoidable, but in practice it is often difficult to overlap
computation and communication to hide their impact. 

Task-based parallelism approaches distributed computing ``from the
other end.'' Instead of splitting the domain into elements, as few
as possible and, each as large as possible (to reduce communication
overhead), one splits the computation into many tasks
of a certain minimum size.
Kaiser et al. \cite{kaiser09} and
Sterling et al. \cite{sterling14} discuss this in greater detail in
the context of programming models for exascale computing, comparing
in particular MPI and a task-based parallelism that is similar to the
one used here. 
The advantages of task-based parallelism are clear --- 
for example load balancing is
much simplified, as the tasks can simply be moved between processes.
However, there is a cost to pay.

Using many small tasks introduces a communication latency delay for
each task, as it has to wait for its input data to be communicated to
the current process. To remedy this, one
employs a \emph{runtime system} (RTS) that keeps track of which task
is waiting for what communication, and that executes tasks as their inputs
become available, preferably in parallel on multiple threads. This scheme
automatically overlaps communication and computation, leading to a
much improved overall efficiency. It is obviously non-trivial to develop 
such a runtime system that executes efficiently on a wide range of
HPC architectures. We thus settled on an existing, proven 
software framework (Charm++~\cite{CharmWebsite}; see below).

A distinguishing feature of SpECTRE is thus its task-based parallelism
strategy that naturally avoids the bottlenecks of globally synchronous
communication.

We are particularly interested in achieving efficiency and scalability to
large (${}\gg O(100\,\mathrm{k})$) core counts. We expect that the defining
features enabling efficient scalability to be
(i) separation of the tasks of communication and computation, so that
they can overlap,
(ii) asynchronous, non-blocking communication so that cores are not
idle, and
(iii) a runtime system to manage task queues, distribute tasks to
cores, and collect timing statistics to inform dynamical load-balancing
decisions. We discuss this in more detail in the next subsection.

When choosing the size of the individual tasks, one needs to strike a
balance between having as many (small) tasks as possible and making
the tasks sufficiently large to overcome the overhead. The relevant
overhead here is the time necessary to create a task, examine its
dependencies, schedule its execution, and move data in and out of CPU
caches. 
On a modern HPC system, this overhead is likely to be in the
range of $1\,\mathrm{\mu s}$ to $10\,\mathrm{\mu s}$; consequently, a
task should run on average for at least that long to be efficient.

\subsection{The Charm++ library}

SpECTRE uses the Charm++ 
library~\cite{CharmWebsite,kale1993charm++,shu1991chare,kale2016parallel}
to implement a task-based parallelism model. 
Among the available task-based libraries we
explored, in our opinion 
Charm++ currently provides the best combination
of robust performance, a diverse set of features, 
good documentation, and a large user base.
The discussion and numerical experiments 
considered throughout this paper use 
version 6.7.0 of the Charm++ library.

The building blocks of a Charm++ application
are distributed objects called {\em chares}. Chares
are ordinary C++ objects with a few special properties:
they inherit from a base class whose 
code is generated by Charm++ at compile time, 
they support creating new instances on remote nodes and
they support a unique kind of
member function called an \emph{entry method}.
When an entry method is invoked a new task is created.
Charm++ takes care of chare distribution and task routing across the
set of available nodes and cores.

SpECTRE's computation is decomposed into a collection of
chares that interact by sending messages to one another.
Two interacting chares may reside on different processors,
and so the sender cannot directly call the receiver's
entry method. Instead, the sender
alerts the Charm++ \emph{runtime system} (RTS)
that a new task needs to be processed. This
request, along with any relevant data, is packaged into a single
message and shipped off to the node on which the
receiving chare resides. The message
is captured by the RTS running on the receiver's node
and placed into a task-pool. The message
remains in the task-pool until the necessary computing
resources become available, at which point the receiver's 
entry method is called and work on the task begins.
A numerical evolution proceeds by asynchronous
message-driven execution~\cite{gursoy2004performance}:
executing tasks create new tasks to perform, and so
on until there are no more
tasks to complete. 

The Charm++ RTS continuously monitors the availability of resources and
keeps track of a list of pending tasks in a task-pool.
Since each instantiated chare resides on a unique
node, local task-pools are maintained 
on each node. When a core becomes available,
the runtime system decides
which message from the task-pool to process
and invokes the entry method specified within the 
message. Control switches back and forth between the RTS's scheduler 
and SpECTRE's code. A message-driven parallel programming model
naturally promotes latency tolerance since there is
never an explicit ``receiving end'' waiting for data
to arrive. Instead, the chare remains dormant  
until all of the necessary remote data has been received.
Meanwhile, other tasks may continue to execute 
on the available computing resources.
In particular, overlap of communication and computation
is much easier to implement than within a traditional MPI framework.
Charm++ also contains features to aid scheduling tasks on GPUs.

In addition to managing task-pools and message routing
between physically separated chares,
the RTS includes numerous features to improve code
performance and maximize processor utilization.
For example, the RTS can optimize
communications between chares on the same node
by exploiting the available shared memory environment.
The RTS also records information about computation
timing statistics and communication patterns for 
measurement-based load balancing. For example,
certain load balancing strategies may attempt to 
cluster frequently communicating chares so that they
reside on nearby physical locations, or dynamically
migrate chares to balance the overall
workload~\cite{zheng2010hierarchical,bhandarkar2001adaptive}. 
We have not yet explored Charm++'s load balancing features.

\subsection{Decomposition of the DG scheme into tasks}

\begin{figure}
\centering
\includegraphics[width=0.9\linewidth]{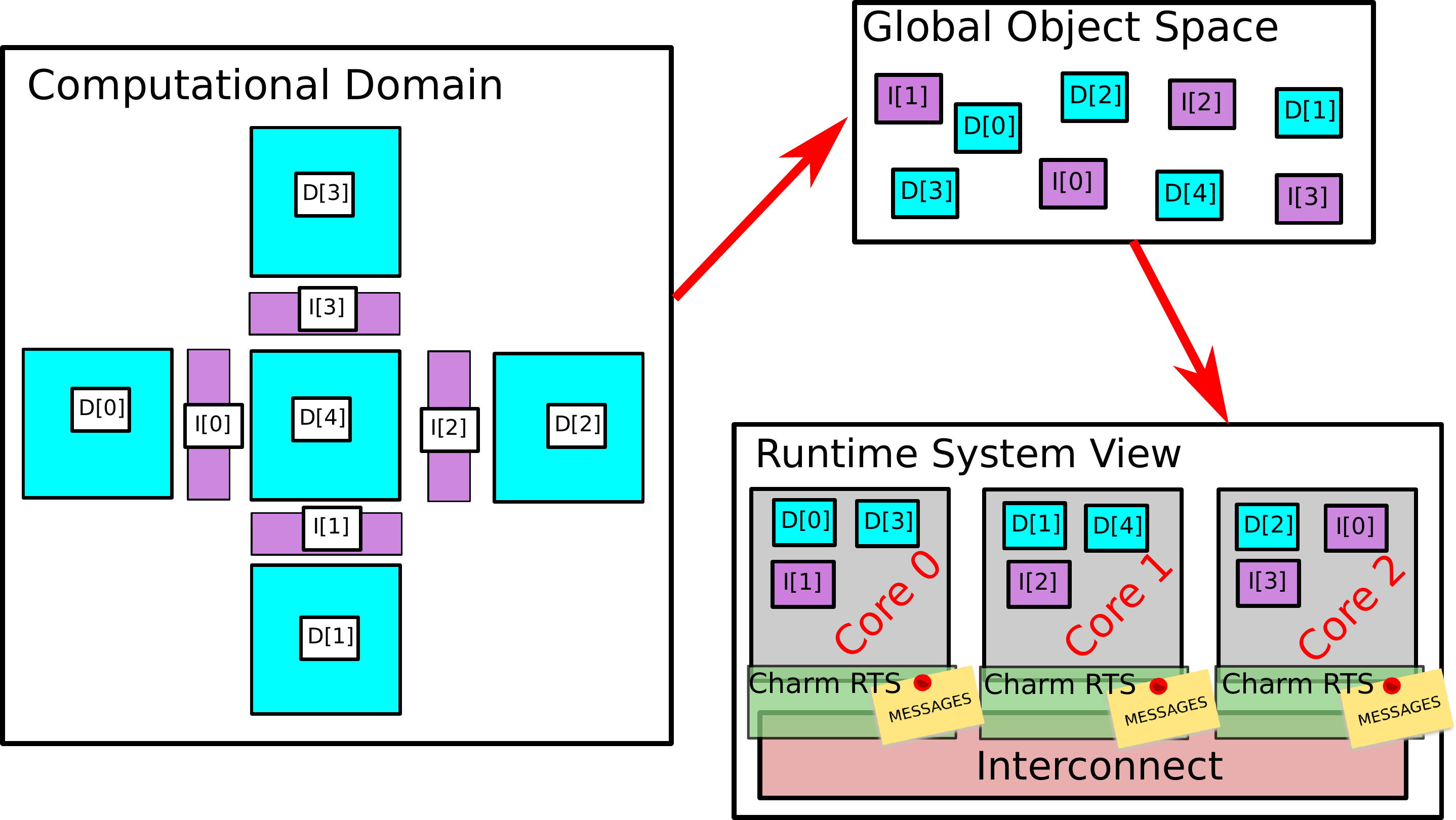}
\caption{Translation of the DG scheme~\eqref{eq:411} 
into a parallelized Charm++ application. First, we
define the ``Computational Domain" as a collection of elements and
interfaces with grid data shown in Fig.~\ref{fig:DGScheme}.
Within the code, each interface (purple) and similarly each element (cyan)
is represented as an element of an array of chares (a chare is a special
Charm++ object). The collection of all chares defines a ``Global Object
Space.'' The code defines tasks that need to be performed by these
objects as well as the communication pattern between objects. Finally, 
at runtime, Charm++ creates interface and element chares on 
available cores to which they are bound. The default distribution
is a round-robin mapping, which is shown in the figure. Tasks that
can be performed on core $0$, for example, are those that are assigned
to elements $0$ and $3$ as well as interface $1$. Task creation, routing,
and scheduling are handled by the Charm++ runtime system (RTS).
}
\label{fig:SpectreDecomposition}
\end{figure} 

The DG algorithm lends itself quite naturally to task-based parallelization. 
The computational domain provides a natural decomposable unit; 
each element and interface is represented as a unique object
in an {\em array of chares}. Figure~\ref{fig:SpectreDecomposition}
depicts the mapping between a computational grid, 
the collection of Charm++ chares needed to represent 
the computational grid, and one possible distribution of these
objects onto a set of three cores with one Charm++ process
running per core. 
In this particular example, five element chares and 
four interface chares are created, each of which contains a portion 
of the grid data illustrated by Fig.~\ref{fig:DGScheme}.
Element and interface chares each define a set 
of tasks (i.e., entry methods) that they will
need to perform.

Each interface chare defines a ``Compute Interface Flux"
task that is invoked by the upper and lower abutting element chares.
Once this task has been called by each neighbor,
the boundary term (right-hand side of Eq.~\ref{eq:411}) is computed and each
element's ``Advance Solution" task is called.

\begin{figure*}
    \centering
    \def\svgwidth{.7 \columnwidth}
    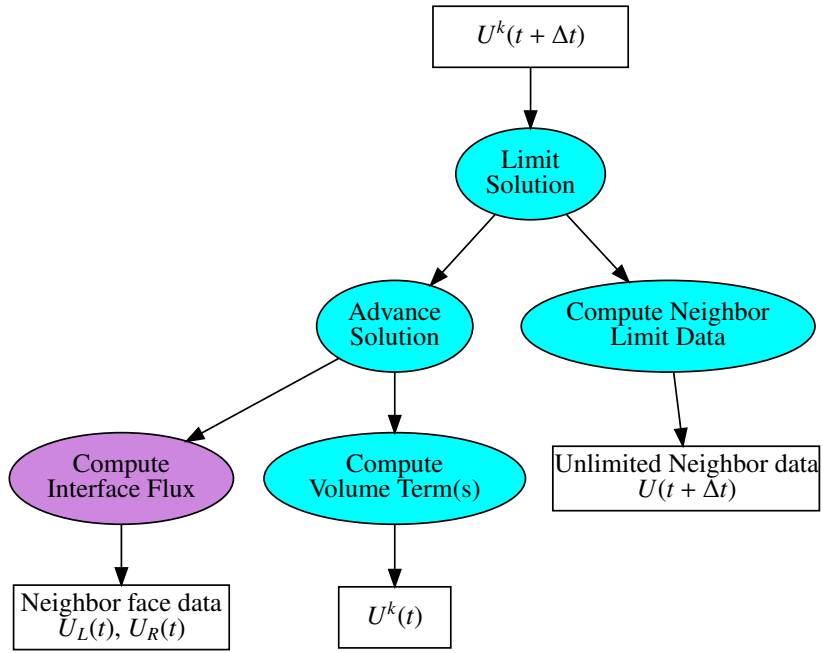
	\caption{Directed acyclic graph showing high-level
task (depicted as ellipses) and data (depicted as rectangles)
dependencies for the DG algorithm advancing the
numerical solution forward in time on a single element.
Tasks are color coded according to
whether they belong to interface (purple) or element (cyan)
chares.
Some tasks are created more frequently
as the number of spatial dimensions increases.
In $d$ spatial dimensions, $2 d$ ``Compute Interface Flux" tasks are 
created during each timestep, for example.
The ``Unlimited Neighbor data" is computed from 
the neighbor's ``Advance solution" data.
\label{fig:task_graph}
}
\end{figure*}

Each element chare defines tasks (i) to compute
the volume terms on the 
left-hand side of Eq.~\eqref{eq:411}, 
(ii) to compute, send and receive 
the data needed by the HRSC limiter (see \S\ref{sec:HRSCLimiters}),
and (iii) to advance the solution forward in time. 
The asynchronous, message-driven parallelism is 
achieved by these ``Charm++-level" tasks. In turn,
each task is composed of traditional routines, 
such as computing primitive variables or equation of state evaluations,
which are executed sequentially
as part of a task's overall workload. 
The task dependencies are specified implicitly by the entry
method calls appearing throughout the code. 
Fig.~\ref{fig:task_graph} represents these dependencies 
when taking a single timestep $t \rightarrow t + \Delta t$
on a single element. This graph represents the minimum
number of tasks for a basic DG scheme, and more complicated 
scenarios involving data observations 
(e.g.~reductions) are
specified as additional tasks, for example.

\subsection{Portability and targeted builds}
\label{sec:charmppBuilds}

The Charm++ library can be installed on a variety of
architectures and interconnect networks. Since the
Charm++
framework naturally decouples low-level parallelization details
from SpECTRE's code, it is straightforward 
to run SpECTRE in different computing environments while 
using machine-specific optimizations provided by
the Charm++ library.
Running on Blue Waters, for example, 
our code passes messages with direct calls to 
Cray's low-level Generic Network Interface (GNI) API
for efficient utilization of the Gemini network~\cite{alverson2010gemini}. 
We have successfully ran SpECTRE
on a variety of other machines supporting a 
variety of communication libraries/infrastructures
including TCP/IP, IB-Verbs (for InfiniBand), and MPI.

We have found the code's performance to be
strongly influenced by how Charm++ is compiled. 
In order to interpret performance metrics 
most clearly (cf.~\S\ref{sec:performance}), we briefly describe
those choices that have the greatest influence on performance.

First, one must select a {\em target architecture}
to build Charm++ for. 
In pure shared-memory environments, such as a desktop,
we always use a ``multicore" target build.
On clusters connected with an InfiniBand network
Charm++ can directly use
InfiniBand's IB-Verbs API, which is enabled
by compiling as a ``verbs-" target build. This is our 
default configuration on systems that support it.
As previously mentioned, 
Blue Waters uses a Gemini network to connect nodes. 
For performance benefits~\cite{sun2012ugni}
Charm++ can directly use Cray's GNI API, which is enabled
by compiling as a ``gemini-gni-crayxe" target build. 

The second choice determines whether or not the Charm++ RTS
should use multiple threads. In keeping with Charm++ terminology,
we shall refer to the no-threading option as a {\em non-SMP build}
and the multi-threaded option as an {\em SMP build}.

A non-SMP build is similar to a pure MPI program. On each core
we launch a single instance of SpECTRE.
Each SpECTRE process is responsible for
executing tasks, managing the task-pool, 
and communicating with all other
processes distributed throughout the supercomputer. By comparison, 
for an SMP build, fewer SpECTRE processes
will be launched than available cores. Each SpECTRE process will 
spawn exactly one thread dedicated for the RTS
and communication, and $N_{\rm work}$ threads to perform tasks. 

Consider, for example, an SMP build and running
on $1$ node with $12$ cores.
We may launch a single process, then spawn a single 
communication thread on core $0$ and $11$ worker threads
on the remaining cores numbered $1$ -- $11$. Another possibility 
would be to launch two processes, each of which spawn 
a single communication and $5$ worker threads (giving $12$ threads
running on all $12$ cores). Typically, if a node has $N$ cores
and we request $N_{\rm com}$ SpECTRE processes
to be launched, there will be $N_{\rm com}$
communication threads (always using $1$ per process) 
bound to $N_{\rm com}$ cores
and 
$N_{\rm work} = N - N_{\rm com}$ 
cores left over to complete tasks. 

With fewer cores available to do computational work, at first sight an
SMP build does not seem competitive. However, as described above,
communication cost is non-negligible. 
One cannot hope to achieve scalability
by assuming that communication cost is near zero, and that one should
``ideally" allocate no resources (no cores) to inter-process
communication.

The advantages of an SMP build include a reduced memory footprint, since
certain data do not need to be duplicated (e.g. global AMR tree
metadata), improved core-to-core connectivity since data can be passed
via pointers instead of making copies, and thus increased scheduling
efficiency since idle cores can pick up work from their neighbors
with less overhead.
In practice, we find the SMP build to be essential when scaling to
large core counts.

%%%%%%%%%%%%%%%%%%%%%%%%%%%%%%%%%%%%%%%%%%%%%%%%%%%%%%%%%%%%%%%%%%%%%%%%%%%%%%%
\section{Performance tests}
\label{sec:performance}
%%%%%%%%%%%%%%%%%%%%%%%%%%%%%%%%%%%%%%%%%%%%%%%%%%%%%%%%%%%%%%%%%%%%%%%%%%%%%%%

As described in \S\ref{sec:charmppBuilds}, Charm++ can be built
in several different ways, and, in our experience, discovering the
``correct" way is often the primary challenge for obtaining good performance
on a new machine.
\S\ref{sec:charmppBuilds}
summarizes those build strategies we have found 
to work best. These settings are used throughout this section.
For all experiments, the Charm++ RTS assigns interface and element chares
to the processors using a round-robin mapping
as depicted in Fig.~\ref{fig:SpectreDecomposition}. 

\subsection{Efficient core usage and asynchronous task execution}
\label{sec:TimeProfile}

Task-based parallelism relies on asynchronous, non-blocking communication 
so that cores can remain active. With Projections, a 
Charm++ analysis tool~\cite{kale2006scaling},
we investigated these features on a single, shared-memory
node of an in-house cluster. We evolved the relativistic MHD system using the
cylindrical blast wave solution described in 
\S\ref{sec:BlastWaveMHD}. The computational grid consisted
of $50\times 50 \times 1$ elements each with
2 GLL nodal points along each dimension in each element
and coupled with an LLF numerical flux. The solution 
was advanced in time using the AB3 stepper. 
After each step we applied the MUSCL slope limiter and
performed a global reduction to compute a pointwise maximum. 

\begin{figure}[htp]
\centering
\begin{subfigure}{0.5\textwidth}
  \includegraphics[width=.95\linewidth]{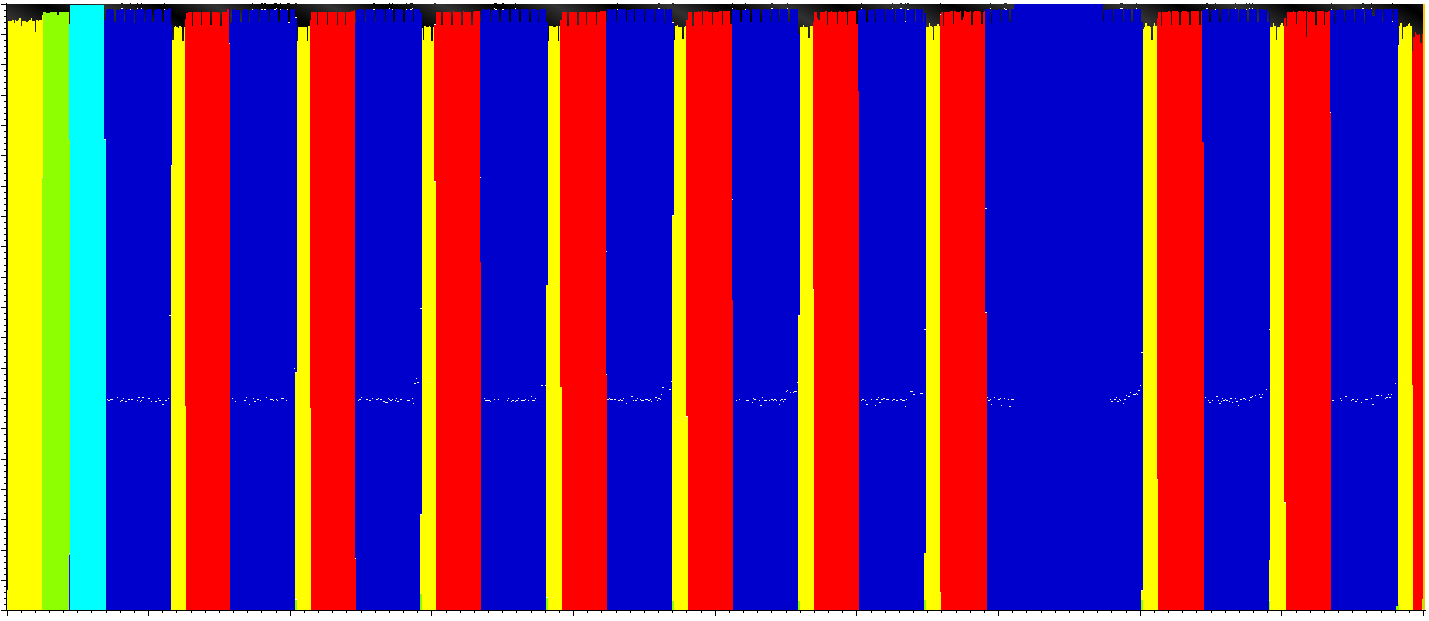}
  \caption{1 core}
  \label{fig:TimeProfileCore1}
\end{subfigure}%
\begin{subfigure}{0.5\textwidth}
  \includegraphics[width=.95\linewidth]{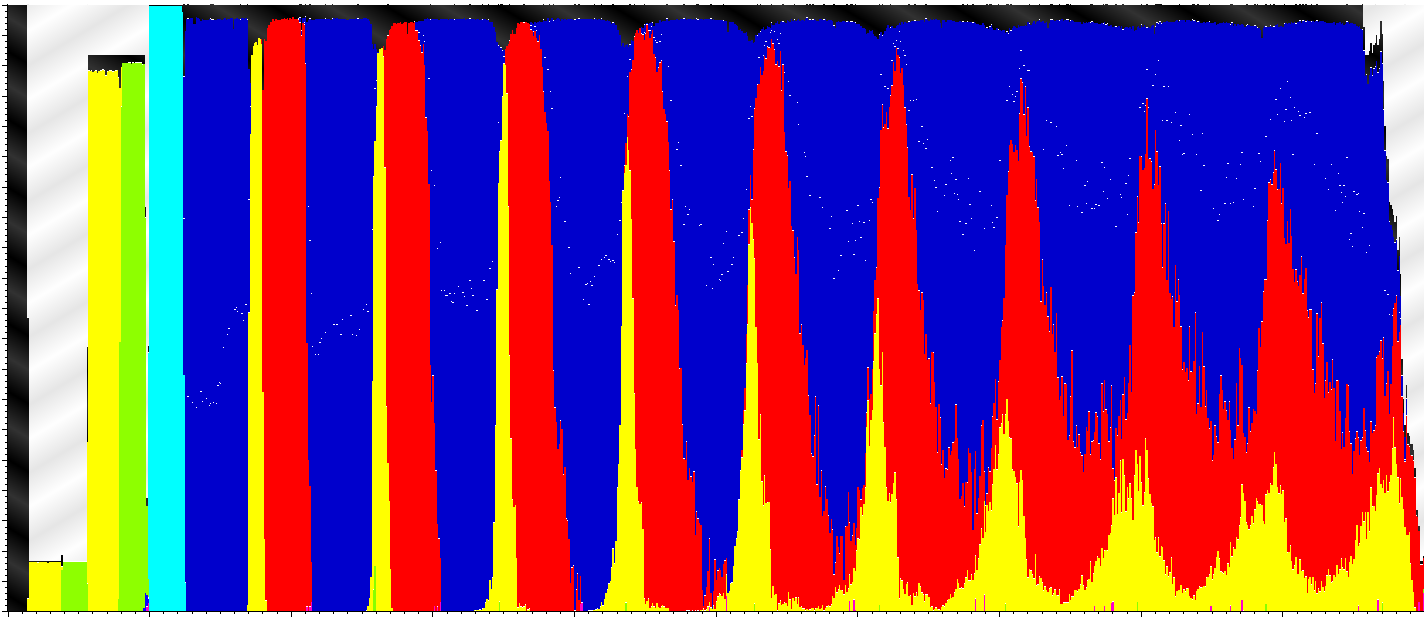}
  \caption{12 cores}
  \label{fig:TimeProfileCore12}
\end{subfigure}
\caption{Total amount of time spent executing each task summed
across all processors in a given time interval. The vertical axis 
shows the combined processor utilization (from $0\%$ to $100\%$)
and the horizontal axis shows the wall time.
The overhead of the Charm++ RTS is depicted as black 
and idle core(s) are depicted as white.
The additional colors show SpECTRE tasks. {\bf Left}: A typical time 
profile graph when running on 1 core, where the simulation is
perfectly synchronized.
{\bf Right}:
A typical time profile graph when running on multiple cores,
in this case 12. Two important features are immediately apparent. 
First, the asynchronous nature of task-based parallelism 
is evident. As the simulation proceeds different parts of 
the computational grid advance at different rates.
Second, all 12 cores remain busy executing
their tasks with essentially zero idle time. 
This would be difficult to achieve within a
traditional parallelism framework employing barriers and waits.}
\label{fig:TimeProfiles}
\end{figure}

We generate time profile graphs for 
a single core (Fig.~\ref{fig:TimeProfileCore1})
and a single node (Fig.~\ref{fig:TimeProfileCore12}) run. Only the most 
computationally intensive tasks are shown. We see that the 
``Compute Interface Flux" (red), ``Compute Volume Terms" (blue),
and ``Limit Solution" (yellow) tasks dominate the simulation's 
cost. When these tasks 
are distributed among the $12$ cores, the asynchronous
communication and execution model manifests itself 
as a nonuniform local simulation time. 
This is clearly seen from the tasks' gradual dispersion
as the simulation proceeds.  Idle cores (shown as white)
only appear at the very end of the simulation after many of the 
chares have finished all of their tasks. In this particular
example, the single core job took about $74$ seconds to complete 
and the single node job took about $6.6$ seconds, demonstrating 
excellent utilization of all $12$ cores and minimal overhead 
of the Charm++ RTS.

\subsection{Strong scaling}
\label{sec:strongscaling}

\begin{figure}[htp]
\centering
  \includegraphics[width=.7\linewidth]{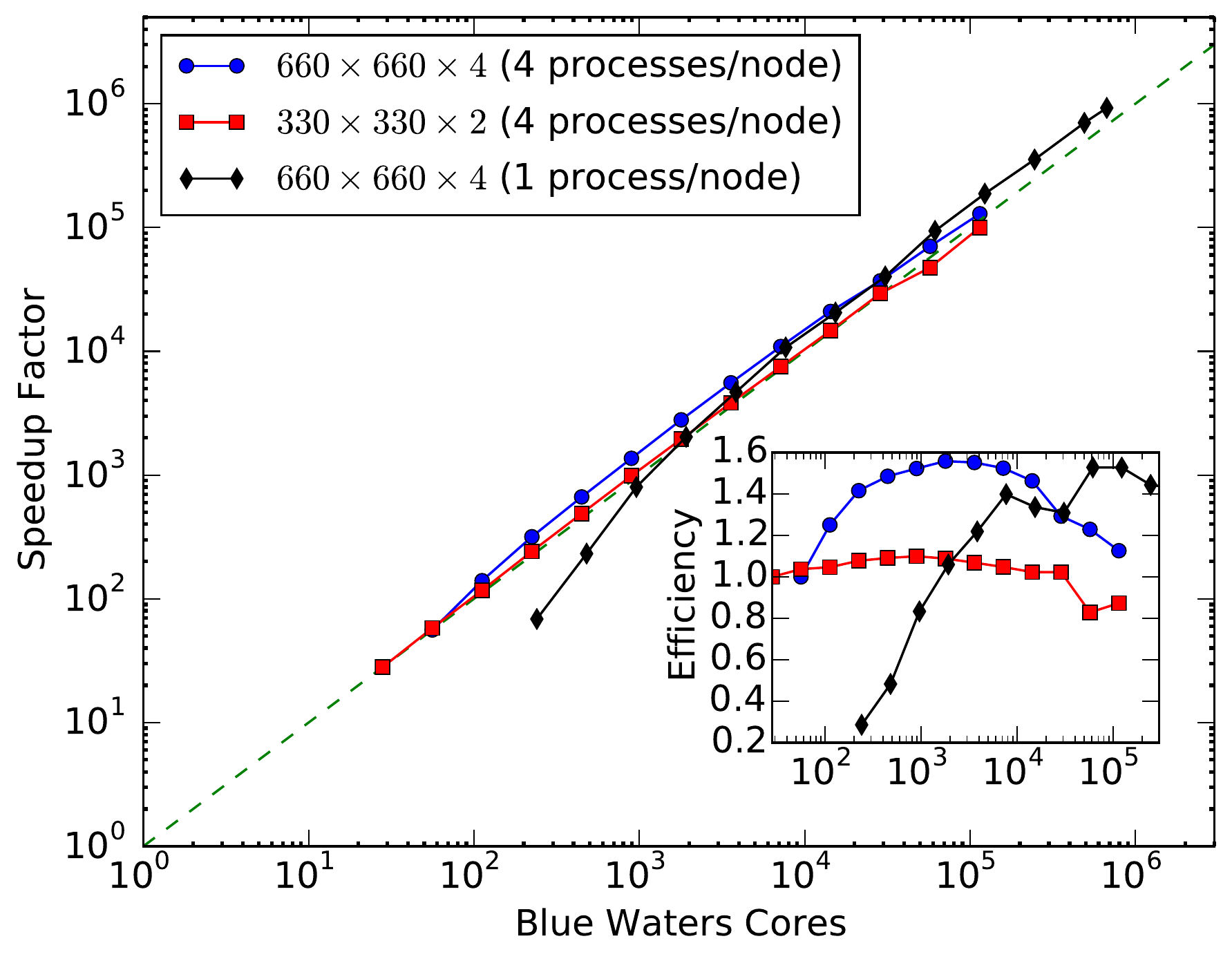}
\caption{
Strong scaling of a SpECTRE evolution of the relativistic MHD system
on Blue Waters for different
grid sizes and number of 
SpECTRE processes per node.
In all cases, each SpECTRE process spawns a single 
Charm++ communication thread responsible for 
passing messages between chares.
For the first two cases using 
$4$ SpECTRE processes per node
(blue circle and red square), speedups and 
efficiencies are computed relative to a hypothetical one core job
whose walltime is defined to be $T_1 = 28 T_{28}$
and where $T_{28}$ is measured directly from a
single-node simulation running $28$ worker threads.
For the cases shown, we measure the walltimes to be
$T_{28} \approx 10.101648$ hours (for the $660\times\, 660\times\, 4$
grid which took $200$ simulation steps) and
$T_{28} \approx 3.433565$ hours 
(for the $330\times\, 330\times\, 2$ grid which took $826$ simulation steps).
Because simulations using 
only $1$ SpECTRE process per node (black diamond) can be especially
inefficient at low core counts (see text), speedups and 
efficiencies are computed relative to the wall time estimated
for a hypothetical single 
core job obtained by a fitting
formula (see text) and found to be $T_1 \approx 712.344370$ 
hours (which took $200$ simulation steps).
Similar scalability results (not shown) have been obtained 
on different machines (e.g.~Stampede, Comet, and SciNet)
and for different evolution systems.
}
\label{fig:StrongScaling-BWs}
\end{figure}

Scaling to a massive number of cores requires sufficient overlap 
of communication and computation so that work is being
done while messages are in flight. This 
obviously requires non-blocking communication, too. We 
demonstrate the effectiveness of our code in
implementing these requirements by considering
a strong scaling test of the relativistic MHD system on Blue Waters
using the test problem described in \S\ref{sec:OTVortex}. 

To assess scalability, we report parallelization efficiency, $E_N$, and 
speedup, $S_N$, given by
\begin{align}
\label{eq:scalability}
E_N = \frac{T_1}{N T_N} \,, \qquad S_N = \frac{T_1}{T_N} \,,
\end{align}
where $N$ is the number of 
``Blue Waters cores'' (BW-cores), 
and $T_1$ and $T_N$ denote the walltime 
respectively using $1$ and $N$ BW-cores.
Every two BW-cores, sometimes 
also called integer scheduling units or 
logical cores, share a single floating point unit (FPU).
Under ideal circumstances, perfect scalability is achieved whenever $E_N = 1$.
In practice, we cannot directly measure $T_1$ since simulations
which run in tens of seconds at large core counts can take weeks or months 
of walltime on a single core -- if the simulation can even fit into
available memory. So we instead estimate $T_1$. Second,
as described below, we sometimes observe values of $E_N$ greater
than one suggesting performance bottlenecks for large problems 
running on small core counts.

Our main result, 
demonstrating efficient usage of Blue Waters
to large core counts, is summarized in Fig.~\ref{fig:StrongScaling-BWs}.
In the figure, 
the number of BW-cores is equal to the number of
worker threads, which gives a modest
performance benefit over launching threads on 
just the floating point units. 
These results have been obtained with an SMP build of Charm++
(see \S\ref{sec:charmppBuilds}). Because of the 
simulation's memory requirements, some configurations could not
be run at low node counts.

The blue circles and red squares shown in Fig.~\ref{fig:StrongScaling-BWs}
represent a similar set of configurations. 
In both cases we have launched $4$ SpECTRE processes 
per node ($2$ processes per socket) each of which spawn 
$1$ communication thread,
leaving $28$ BW-cores available on each node to complete tasks.
The total number of BW-cores quoted for these
cases is $28$ times the number of nodes.
We ran a sequence of jobs from $1$ node ($28$ BW-cores) to 
$4,096$ nodes ($114,688$ BW-cores). With a computational
grid composed of $330 \times 330 \times 2$ elements,
good scalability is observed up to the highest core count where, on average,
each core is responsible for only $\approx 2.8$ elements.
Repeating the experiment after refining the  
grid to $660 \times 660 \times 4$ elements
shows similar scalability, 
albeit with reduced performance
at low core counts (see discussion below).
In both cases, we compute measures of 
scaling performance~\ref{eq:scalability}
relative to a hypothetical one core job
whose walltime is defined to be $T_1 = 28 T_{28}$
and where $T_{28}$ is measured directly from a
single-node simulation. Note that our definition for $T_1$ 
estimates its true value
by assuming perfect intra-node scaling.

Beyond $4,096$ nodes our simulations exhausted
the system's available memory pages. A possible
explanation could be that with more messages in flight,
more pages are necessary to accommodate message buffers.
We work around this problem
by launching fewer processes 
per node. For these runs,
which used just one 
SpECTRE process (hence just one communication thread)
per node,  
we also left one core free for system processes, leaving 
$30$ threads per node for work. The black diamonds 
shown in Fig.~\ref{fig:StrongScaling-BWs} represent 
data from this configuration, which continues to 
show excellent scaling up to the
machine's \emph{entire} 
pool of $22,380$ available nodes ($671,400$ cores)
despite an average of just $\approx 2.8$ elements per core.
Because simulations using 
only $1$ SpECTRE process per node can be especially
inefficient at low core counts (see below), speedups and 
efficiencies are computed relative to the wall time estimated
for a hypothetical single core job obtained by fitting
a portion (from $3,840$ to $491,520$ BW-cores)
of the walltime data to a straight line and evaluating this fit
at one core.

In all cases considered, we find that whenever the
number of chares per process becomes too large,
our code becomes inefficient. This could be because we
overwhelm the Charm++ RTS with too many messages; the Charm++
documentation and mailing lists mention this as a potential problem.
As a future performance improvement we may consider modifying the code to 
pack more elements into a chare, thereby reducing the number of chares
per process.

As a final measure of the code's performance, we also record the 
CPU time per timestep per total number of simulation 
grid points (e.g. the first sequence of simulations
uses $660\times660\times4\times 2^3$ points, about 
$14$ million). Here the CPU time is computed as the 
wall time (which is directly measured) multiplied by the number 
of floating point units being utilized by the worker 
threads (two worker threads share a single floating
point unit). At large core counts,
jobs using $4$ processes
per node behave similarly, with e.g. $\approx 110\, \mathrm{\mu s}$
needed per step per grid point when running on 
$896$ cores. Jobs using $1$ process per node
require $\approx 300\, \mathrm{\mu s}$ when 
running on $61,440$ cores.

\subsection{Weak scaling}

Figure~\ref{fig:WeakScaling} demonstrates the code's scalability on
the San Diego Supercomputer Center's machine Comet.
Each compute node features two Intel Xeon E5-2680v3 2.5 GHz 
chips, each equipped with $12$ cores, for 
a total of $24$ cores per node. Nodes are 
connected by an InfiniBand interconnect.

{Having just shown strong scalability, we now report on a weak scaling 
experiment on Comet whereby the number of elements per core is held fixed 
as the number of nodes is increased from 
$1$ ($24$ cores) to $72$ ($1,728$ cores), the maximum 
allowable amount on this machine. Computational grids with
heavy ($8$ elements per 
core with degree $N=3$ basis functions) and 
light ($1$ element per core with degree $N=1$ basis functions) workloads show 
similar trends up to this machine's largest allowable core 
allocation. Non-SMP jobs use $24$ processes 
per node and SMP jobs launch $4$ communication threads per node 
leaving $20$ worker cores available. 

\begin{figure}[htp]
\centering
  \includegraphics[width=.6\linewidth]{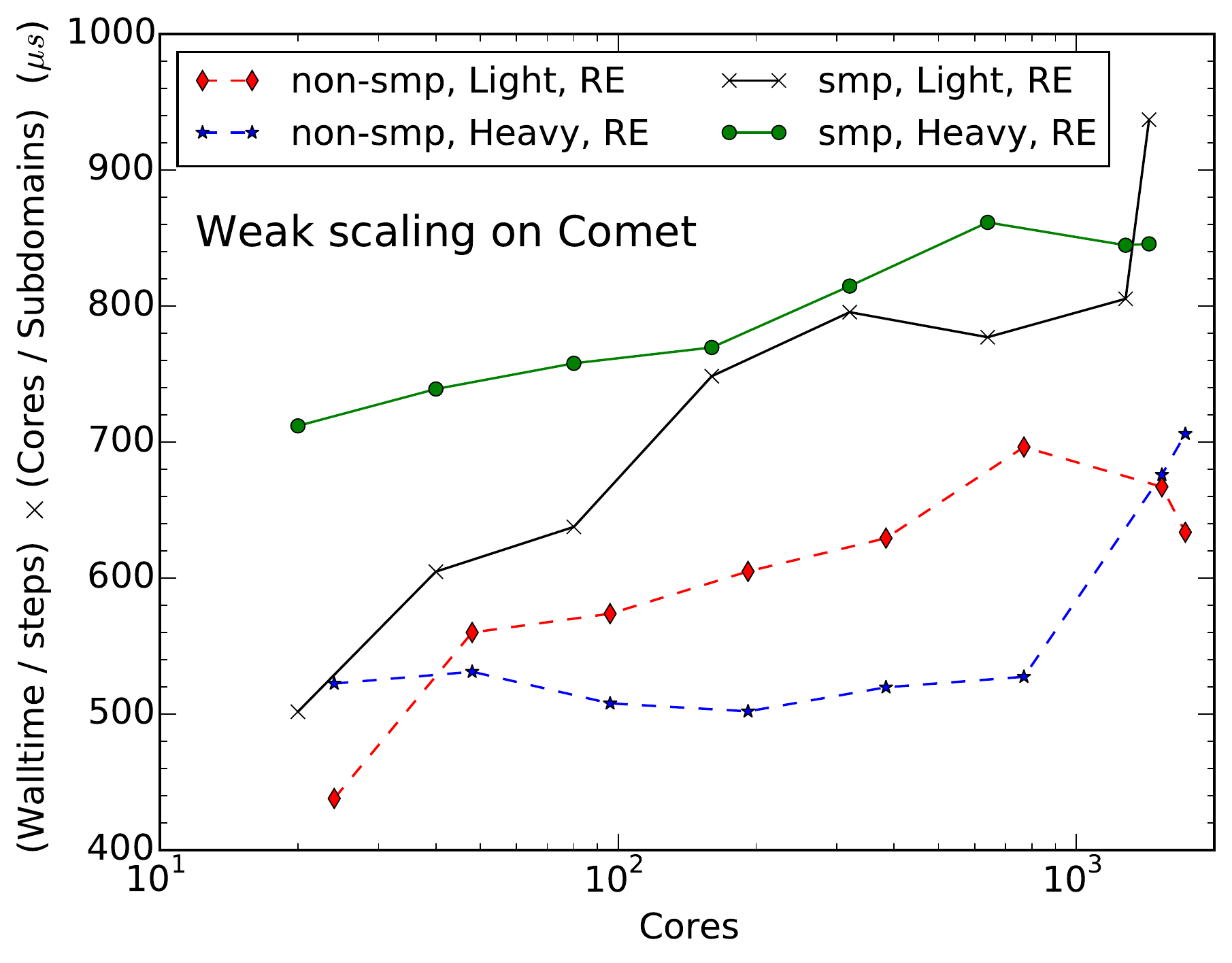}
\caption{Weak scaling of a SpECTRE evolution of
the relativistic hydrodynamics system on Comet with (SMP)
and without (non-SMP) multithreading. 
The problem setup is identical to the smooth flow 
convergence test described in \S\ref{sec:RESmooth}.
We consider computational grids
with a ``heavy" and ``light" amount of work, and
report a CPU time (lower is better) which has 
been scaled by the number of elements to facilitate
the comparison of different grid resolutions.
Similar scalability results (not shown) have been obtained 
on different machines (e.g.~Stampede and SciNet)
and for different evolution systems.
}
\label{fig:WeakScaling}
\end{figure}

%%%%%%%%%%%%%%%%%%%%%%%%%%%%%%%%%%%%%%%%%%%%%%%%%%%%%%%%%%%%%%%%%%%%%%%%%%%%%%%
\section{Benchmark tests}
\label{sec:tests}
%%%%%%%%%%%%%%%%%%%%%%%%%%%%%%%%%%%%%%%%%%%%%%%%%%%%%%%%%%%%%%%%%%%%%%%%%%%%%%%

In this 
section we present a number of useful benchmark tests in one and three spatial
dimensions. All of our tests report on simulations of the 
Newtonian Euler (NE), Relativistic Euler (RE), and 
Relativistic MHD (RMHD) systems.
Our goal is both to demonstrate that the numerical scheme
works as expected and to perform
comparisons with other codes or exact solutions. Except for a few special
instances, we shall largely avoid comparing the accuracy and efficiency 
of numerical schemes that differ by choice of, for example, the 
numerical flux. We shall sometimes use the shorthand tag
\{Numerical Flux\}-\{Slope Limiter\}-\{GLL points per dimension per element\} 
to refer to a particular numerical scheme. For example,
``Roe-MRS50-3" would denote a simulation using 
a Roe numerical flux 
with $3$ GLL points along each dimension
(basis functions of degree $2$ polynomials) 
and an MRS limiter with $\tilde \alpha = 50$.

Unless stated otherwise, we report the numerical error as an $L_1$ norm,
\begin{align} \label{eq:L1}
\| u - v\|_{L_1} = \sum_{i=0}^{n-1} \int_{\Omega} \left| u_i - v_i \right| dV\,.
\end{align}
Here $u$ and $v$ are the analytic
and numerical solution vectors. The integral is computed using a 
Gauss-Legendre-Lobatto quadrature rule
(see Algorithm $25$ from~\cite{kopriva2009implementing}).

\subsection{One-dimensional tests}

\subsubsection{Smooth flow (RE) and comparison to Bugner et al.}
\label{sec:bugner}

Our first test compares to the code of Bugner et al~\cite{Bugner:2015gqa}
by replicating their ``smooth sine wave" 
special relativistic hydrodynamics test (problem $51$ in Bugner).
For this one-dimensional test, their code is algorithmically
equivalent to ours if their mass matrix is computed with the same
Gauss-Lobatto technique described in \S\ref{sec:algorithm} so that
it is diagonal.

To repeat their test, we consider the $1$-dimensional analog of
Eq.~\eqref{eq:smooth_flow_re} using a setup identical to theirs and
described in the caption of Table~\ref{tab:Bugner_comparison}.  Our
errors are quoted in Table~\ref{tab:Bugner_comparison}. Comparing to
their $L_1$-errors~\cite{Bugner:Email:2016,Bugner:2015gqa}
we find excellent agreement with their code, with relative differences
smaller than $5 \times 10^{-12}$. These results constitute a strong,
independent sanity check for both codes.
More details can be found in Appendix B of Ref.~\cite{Bugner:2015gqa}.

For the smooth solution considered here, 
Fig.~\ref{fig:RelEuler1DConvergence} shows the expected spectral (exponential)
convergence in the approximation. That is, for a fixed number of uniform
elements, as we increase the polynomial degree $N$, the error~\eqref{eq:L1}
typically
decreases like $\sim \exp\left( - c N \right)$ for some 
positive constant $c$. This property of a DG scheme is observed
in all tests whose solution is smooth enough.

\begin{figure}
\centering
\includegraphics[width=0.99\linewidth]{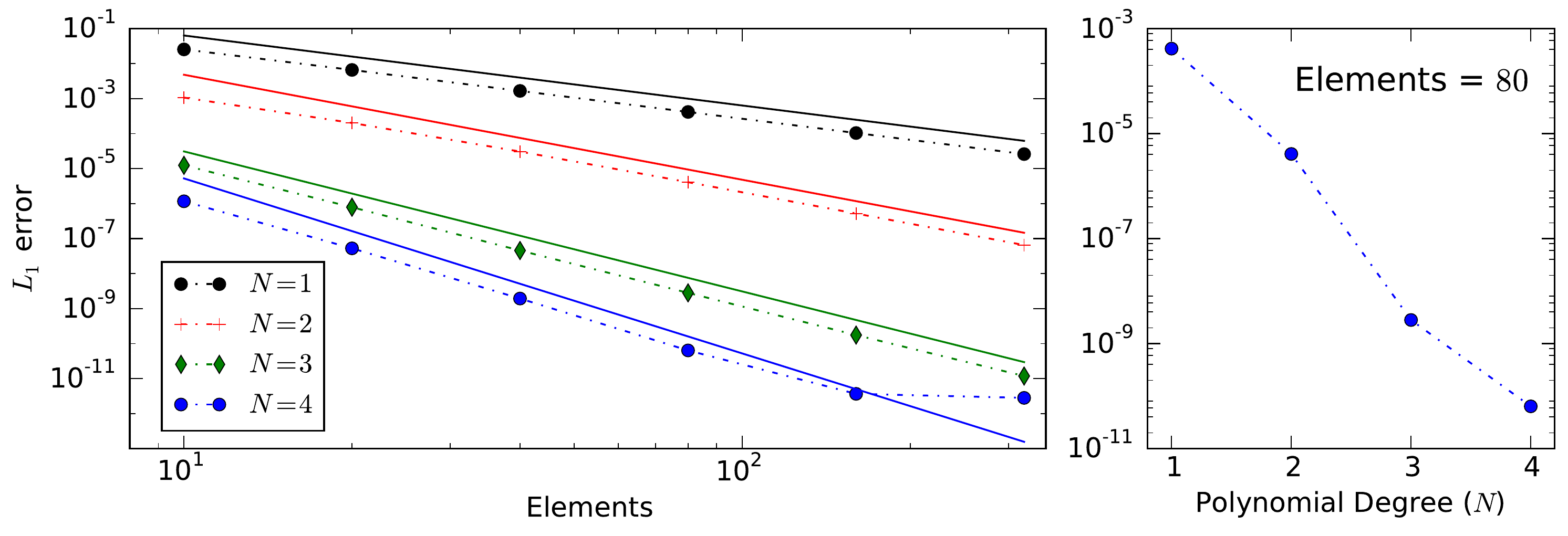}
\caption{Convergence of the numerical solution
for a sequence of simulations 
performed with
the relativistic hydrodynamics
system and described further 
in \S\ref{sec:bugner}. We consider convergence
by increasing the number of elements (left) and 
polynomial degree of the numerical 
approximation (right). These error profiles
are typical of our DG scheme when the solution is smooth.
In this case the DG scheme achieves exponential
convergence in the approximation error as the polynomial
degree $N$ is increased. For a fixed value of $N$,
the approximation error decreases with a power law (dashed line)
at a rate which closely matches the
expected rate of $-\left(N+1\right)$ (solid line). 
The data plotted here is also
given in  Table~\ref{tab:Bugner_comparison}.
}
\label{fig:RelEuler1DConvergence}
\end{figure}

\subsubsection{Riemann problems (NE)}
\label{sec:RP_NE1d}

\begin{table*}
\setlength{\tabcolsep}{6pt}
\centering
\begin{tabular}{c | c c  c  c  c  c  c c c}
Name & System & $\rho_L$ & $u_L$ & $P_L$ & $\rho_R$ & $u_R$ & $P_R$ & $\Gamma$ & $x_d$ \\
\hline
Sod & Euler  & $1.0$ & $0.0$ & $1.0$ & $0.125$ & $0.0$ & $0.1$ & $1.4$ & $0.0$ \\
Shock tube 1 & SR Euler & $10.0$ & $0.0$ & $13.33$ & $1.0$ & $0.0$ & $10^{-8}$ & $5/3$ & $0.5$ \\
Shock tube 2 & SR Euler & $1.0$ & $0.0$ & $1000.0$ & $1.0$ & $0.0$ & $10^{-2}$ & $5/3$ & $0.5$
\end{tabular}
\caption{Initial data for Riemann problem tests.
In all cases, the fluid is taken to be an ideal gas with an adiabatic
index $\Gamma$ and the initial discontinuity located at $x_d$.
\label{tab:Riemann_Problems}
}
\end{table*}

Riemann problems comprise an important
class of tests for a hydrodynamics code. 
These one-dimensional problems 
specify constant initial data for the left, $U_{\rm L}$, 
and right, $U_{\rm R}$, states
\begin{equation} \label{eq:RP}
U =
\begin{cases}
U_{\rm L}, & \text{for } x \leq x_d \\
U_{\rm R}, & \text{for } x > x_d 
\end{cases}
\end{equation}
on either side of a discontinuous interface, which we have 
taken to be located at $x=x_d$. Depending on the 
values for the left and right states, 
this initial data may evolve into some
combination of a contact discontinuity, 
shock, and rarefaction wave. The specific 
wave-like configuration is computable from 
the initial data~\cite{gheller1997high}, which in turn permits 
the computation of an exact solution~\cite{lora2013exact}.

For the Newtonian Euler system we shall consider
the standard Sod shock tube test
whose initial data is given in Table.~\ref{tab:Riemann_Problems}.
Our purpose is to verify basic properties of the 
numerical scheme and its ability to resolve sharp features.
\ref{app:sod} explores the strengths 
and weaknesses of different limiters and numerical fluxes
and provides additional details.
In all cases, our numerical scheme
exhibits a convergence rate of $\approx 1$ (see Table.~\ref{tab:NE_sod}), as
expected for discontinuous solutions of this type.
A visual inspection (see Figure~\ref{fig:Sod1D_CompareLimiter}) 
of the solution reveals, broadly
speaking, that the Roe flux outperforms
all three of its competitors while 
the Local Lax Friedrichs flux typically performs the worst.

\subsubsection{Riemann problems (RE)}
\label{sec:RP_RE1d}

We now consider two Riemann problems~\eqref{eq:RP}
in the context of special relativistic hydrodynamics,
again comparing our numerical solution to
an exact solution~\cite{rezzolla2001improved}. 
These are standard tests frequently used to verify
relativistic hydrodynamic codes (e.g.~\cite{lucas2004assessment,zhang2006ram}). 

The tests we consider here 
are more challenging than the Sod shock tube problem
of \S\ref{sec:RP_NE1d}. Numerical errors can
lead to situations where
the density becomes negative or there is no solution
to the equations expressing the primitive variables in
terms of the conserved ones.
To remedy these problems,
in low-density regions we follow the prescription described in 
the ``Atmosphere treatment" appendix of Ref.~\cite{Muhlberger2014},
which guarantees a solution exists with physically reasonable 
bounds on the primitive variables.

The first Riemann problem we consider is ``shock tube 1" whose
initial data is given in Table.~\ref{tab:Riemann_Problems}. The solution
develops a left-moving rarefaction wave, a right-moving contact wave, and
a right-moving shock wave. The fluid's velocity reaches a mildly-relativistic
maximum of $\approx 0.7$. 
The expected order of convergence (see Table~\ref{tab:RE_sod}) 
and good resolution of the discontinuous features
(see Figure~\ref{fig:RE1d_Shocks}) are observed over 
a wide range of numerical fluxes and limiters.  A few of the 
$\approx 50$ tests we ran on this problem failed.
Among the four possible fluxes, the Marquina flux
failed most often, especially when paired with the MRS limiter.
The MRS limiter failed more often
whenever using $\alpha \geq 50$ and $N \geq 2$.

The next Riemann problem we consider is ``shock tube 2" whose
initial data is given in Table.~\ref{tab:Riemann_Problems}. The solution
develops the same wave structure as ``shock tube 1", but now the 
fluid's velocity reaches into the relativistic regime. The shock's speed,
for example, is $\approx 0.986$ corresponding to a
Lorentz factor of about $6$~\cite{lucas2004assessment}.
An extremely thin ``blast wave" of high 
density develops between the contact discontinuity and the shock. 
Because of the blast wave's narrow width, $\approx 0.01$
at $t=0.4$~\cite{zhang2006ram}, most codes have difficulty 
resolving this solution except at very 
high grid resolutions~\cite{lucas2004assessment,zhang2006ram}.
At high resolutions our DG scheme is able to capture the 
features of this solution (Figure~\ref{fig:RE1d_Shocks})
and we observe convergence a bit lower than first order 
(see Table~\ref{tab:RE_sod}),
which has also been observed in finite volume codes running 
this test~\cite{zhang2006ram}. 
All $\approx 50$ numerical schemes we tried 
were stable for this test case.

\begin{figure}
\centering
\includegraphics[width=0.95\linewidth]{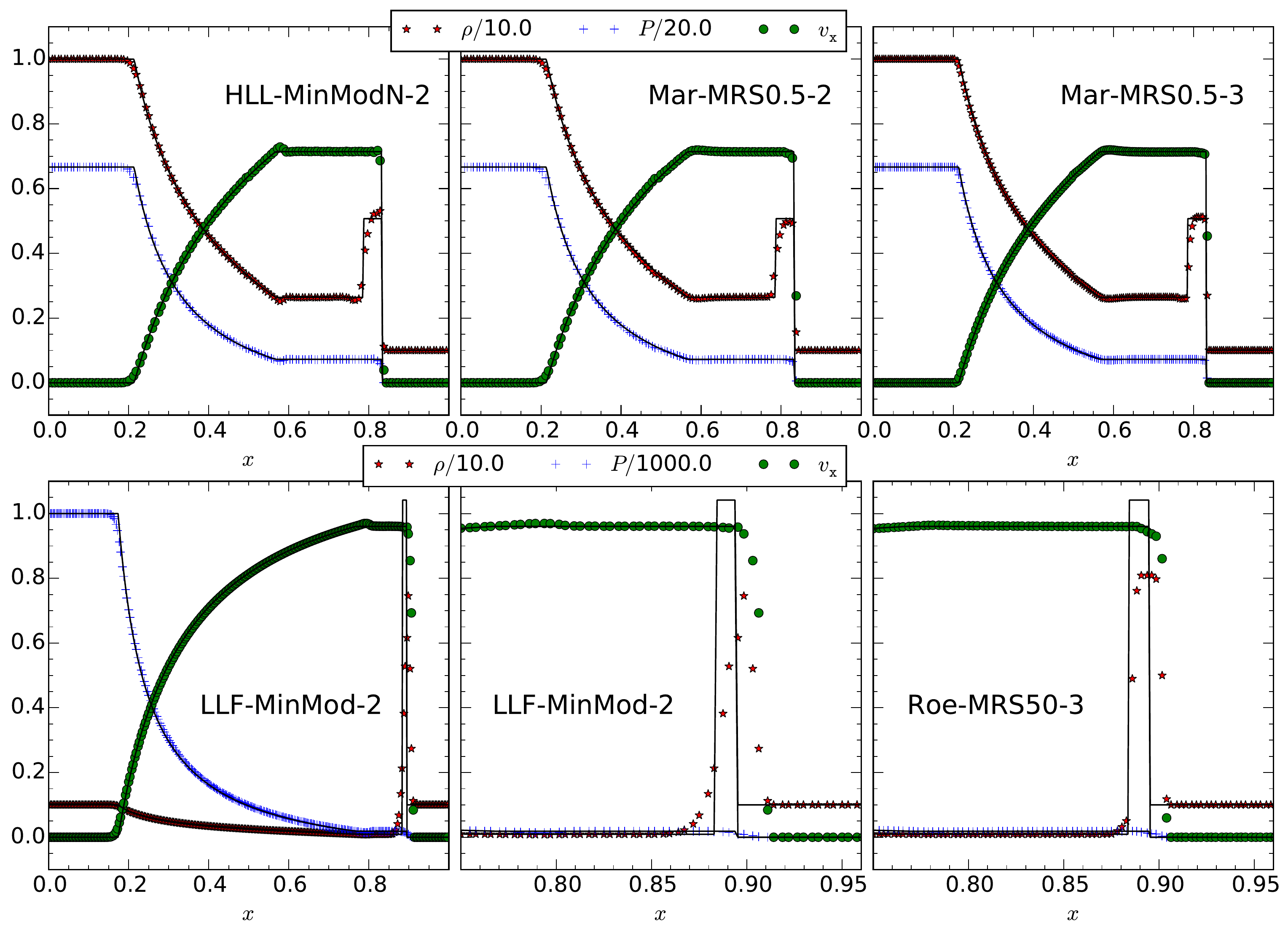}
\caption{Comparison of the special relativistic 
shock tube problems computed with 
different limiters and numerical fluxes. 
The setup is described 
in the caption of Table.~\ref{tab:RE_sod}
and in both cases we evolve on 
 $n_x = 640$ uniform DG elements
to $T=0.4$.
{\bf Top row}: Solution to 
shock tube 1 problem. 
{\bf Bottom row}: Solution to the highly relativistic 
shock tube 2 problem.
The middle and right panels zoom-in on
the sharp blast wave feature 
that makes this a particularly difficult problem.
}
\label{fig:RE1d_Shocks}
\end{figure} 

\subsubsection{Riemann problems (RMHD)}
\label{sec:testrmhd}

To test the ability of our code to handle shocks when using the
relativistic equations of magnetohydrodynamics (RMHD), we first
consider a standard suite of 1D RMHD shock tests due to
Komissarov~\cite{Komissarov1999}. These tests are all performed in
Minkowski spacetime. The initial configuration consists of a left
state $U^L$ for $x<0$ and a right state $U^R$ for $x>0$, each with
constant fluid variables. The initial conditions for the different
tests are listed in Table~\ref{tab:komissarov}. All tests use an ideal
gas equation of state $P=(\gamma-1)\epsilon \rho$, with $\gamma=4/3$.
We note that although the numerical domain in these tests is
one-dimensional, the evolved variables (velocity, magnetic field) are
three-dimensional. For all tests, we use a resolution $\Delta x =
0.01$, and a third-order Runge-Kutta time stepper.

\begin{table*}
\caption{\label{tab:komissarov}Initial data for the RMHD shock tests based on Komissarov~\cite{Komissarov1999},
and the more relativistic two-stream collision test from Balsara~\cite{Balsara:2001} 
(Test 4 in Table I of~\cite{Balsara:2001}, Collision 2 here).}
\begin{tabular}{lll}
\hline
Test & Initial state for $x<0$ & Initial state for $x>0$ \\
\hline
Fast shock & $\rho=1$, $P=1$  &  $\rho=25.48$, $P=367.5$  \\
($t_\text{final}=2.5$) & $u_i=(25,0,0)$, $B^i=(20,25.02,0)$ & $u_i=(1.091,0.3923,0)$, $B^i=(20,49,0)$ \\
\hline
Slow shock & $\rho=1$, $P=10$&  $\rho=3.323$, $P=55.36$ \\
($t_\text{final}=2.0$)  & $u_i=(1.53,0,0)$, $B^i=(10,18.28,0)$  & $u_i=(0.9571,-0.6822,0)$, $B^i=(10,14.49,0)$\\
\hline
Switch-off & $\rho=0.1$, $P=1$&  $\rho=0.562$, $P=10$ \\
($t_\text{final}=1.0$)  & $u_i=(-2,0,0)$, $B^i=(2,0,0)$  & $u_i=(-0.212,-0.590,0)$, $B^i=(2,4.71,0)$\\
\hline
Switch-on & $\rho=0.00178$, $P=0.1$&  $\rho=0.01$, $P=1$ \\
($t_\text{final}=2.0$)  & $u_i=(-0.765,-1.386,0)$, $B^i=(1,1.022,0)$  & $u_i=(0,0,0)$, $B^i=(1,0,0)$\\
\hline
Shock tube 1 & $\rho=1$, $P=1000$&  $\rho=0.1$, $P=1$ \\
($t_\text{final}=1.0$)  & $u_i=(0,0,0)$, $B^i=(1,0,0)$  & $u_i=(0,0,0)$, $B^i=(1,0,0)$\\
\hline
Shock tube 2 & $\rho=1$, $P=30$&  $\rho=0.1$, $P=1$ \\
($t_\text{final}=1.0$)  & $u_i=(0,0,0)$, $B^i=(0,20,0)$  & $u_i=(0,0,0)$, $B^i=(0,0,0)$\\
\hline
Collision & $\rho=1$, $P=1$&  $\rho=1$, $P=1$ \\
($t_\text{final}=1.22$)  & $u_i=(5,0,0)$, $B^i=(10,10,0)$  & $u_i=(-5,0,0)$, $B^i=(10,-10,0)$\\
\hline
Collision 2 & $\rho=1$, $P=.1$&  $\rho=1$, $P=.1$ \\
($t_\text{final}=0.4$)  & $u_i=(22.344,0,0)$, $B^i=(10,7,7)$  & $u_i=(-22.344,0,0)$, $B^i=(10,-7,-7)$\\
\end{tabular}
\end{table*}

\begin{figure}
\centering
\includegraphics[width=0.95\linewidth]{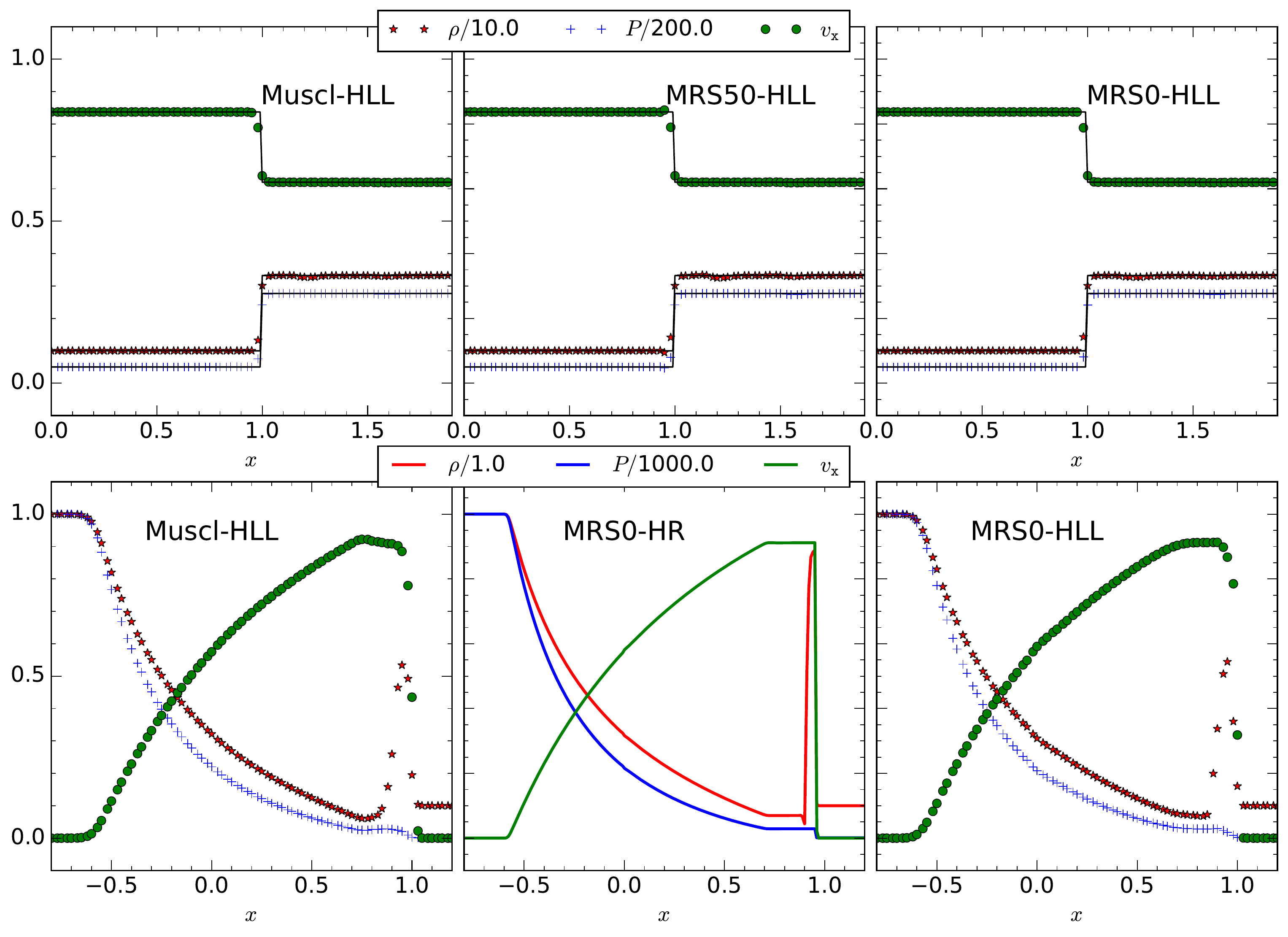}
\caption{Comparison of the special relativistic MHD
shock tube problems computed with 
different limiters. We show the SlowShock (top) and
ShockTube1 (bottom) tests described in Table~\ref{tab:komissarov}.
We solve the test on uniform elements with grid spacing 
$\Delta x = 0.01$ and $2$ GLL points per DG element. For the
SlowShock test, the analytical solution is shown as a thin black line.
For the ShockTube1 test, we compare the numerical solution 
to a high resolution (``MRS0-HR") simulation
($\Delta x =0.001$) in the middle panel.
}
\label{fig:RMHD1d_Shocks}
\end{figure} 

Results for the SlowShock test and the ShockTube1 test are shown in
Fig.~\ref{fig:RMHD1d_Shocks}.  The results of these tests are similar
to those some of us obtained with a finite volume, relativistic MHD
code in~\cite{Muhlberger2014}.  The only exceptions are when using the
MRS limiter with more than $2$ GLL points per element, and for the
FastShock test. For high-order elements, the MRS limiter with
$\alpha>100$ does not appear to be dissipative enough to stably evolve
shocks. 

The FastShock test, on the other hand, only converges to the correct
solution if cells in which the conserved variables have evolved towards unphysical
values are flagged as troubled, and treated with any of the limiters discussed
in this work. If conserved variable fixing is applied point-wise to a cell
in which no limiting has occurred, on the other hand, the shock front
propagates at a speed $2\%-5\%$ slower than the expected value.
The exact propagation speed depends on the choice of
limiter, but converges to the wrong value for our standard
limiters. This is likely to be a consequence of the fact that fixing the 
conserved variables is not a conservative operation, while limiting
conserves the integrated value of the evolved variables within a cell.
We note,
however, that the FastShock test has a Lorentz factor $\Gamma \sim 25$
and a plasma parameter $\beta = 2P/b^2\sim 0.002$. This is a regime in
which global simulations of binary mergers and accretion disks no
longer try to accurately evolve the GRMHD system of equations, and
instead impose an upper bound on $\Gamma$ and/or a lower bound on the
density (artificial atmosphere) [see e.g.~\cite{Muhlberger2014}].

We also perform the more demanding two-stream collision
test from Balsara~\cite{Balsara:2001}. That test uses an ideal gas
equation of state with $\gamma=5/3$, and involves Lorentz factor
$\Gamma \geq 22$. For this test and all other configurations from
Komissarov~\cite{Komissarov1999}, the results are similar to those
obtained using finite volume methods 
(e.g.~\cite{Komissarov1999,Balsara:2001,Muhlberger2014}). Both
the MRS and Muscl limiters robustly evolve these 1D tests, with the
MRS limiter typically resolving shock fronts more accurately, but the
Muscl limiter avoiding the creation of sharp, unphysical features
more robustly.

\subsection{Three-dimensional tests}

\subsubsection{Isentropic vortex (NE)}

The isentropic vortex
solution~\footnote{The vortex solution
presented in Ref.~\cite{yee1999low} and its generalization
in Ref.~\cite{Hesthaven2008} contain typos. 
A corrected expression can be found in the
online errata for Ref.~\cite{Hesthaven2008}.} 
, which was first described by Jiang and Shu~\cite{Jiang1996202},
is given by
\begin{align} \label{eq:isen_vortex}
\rho  = \left(1 - \frac{(\gamma - 1) \beta^2}{8 \gamma \pi^2} e^{1-r^2} \right)^{1/(\gamma-1)} \,,
\quad v_x  = U - \tilde{y} \frac{\beta}{2 \pi} e^{(1-r^2)/2} \,,
\quad v_y = V + \tilde{x} \frac{\beta}{2 \pi} e^{(1-r^2)/2} \,,
\quad v_z = W  \,, 
\end{align}
where $\tilde{x} = x- X_0 - U t$, $\tilde{y} = y-Y_0-V t$, and $r^2 =
\tilde{x}^2 + \tilde{y}^2$. 
This is essentially a two-dimensional problem that we solve 
on a three-dimensional computational grid.
The constants $U$, $V$, and $W$ specify
the velocity of the mean fluid flow, $X_0$ and $Y_0$ locate the vortex
center at $t=0$, $\beta$ sets the vortex strength, and $\gamma$ is the
polytropic index for the adiabatic equation of state $P =
\rho^\gamma$.  Our numerical experiment, fully described in the
caption of table~\ref{tab:IsenVortexConvergence}, varies both the
polynomial degree and number of elements. The resulting numerical
errors, quoted in Table~\ref{tab:IsenVortexConvergence}, exhibit the
expected convergence rate for a fixed polynomial degree $N$ and
spectral convergence with $N$.

\subsubsection{Smooth flow (RE)}
\label{sec:RESmooth}

We consider a relativistic generalization of the periodic, smooth
flow solution of Xu et al.~\cite{xu2009hierarchical} given by
\begin{align} \label{eq:smooth_flow_re}
\rho = 1 + A \sin(u) \,, 
\quad \epsilon = \frac{1}{\left(\gamma - 1 \right)\rho} \,,
\quad v_x = U \,, 
\quad v_y = V\,, 
\quad v_z = W \,,
\end{align}
where $u = k \left[x + y + z - (v_x + v_y + v_z) t\right]$.
The constants $U$, $V$, and $W$ specify the velocity of the mean fluid flow,
the constant $A \leq 1$ describes the size of the fluid's oscillation
about unity, and $k$ is the wave number. 
This choice of specific internal energy is arranged to give a
constant pressure, $P = \left( \gamma - 1 \right) \rho \epsilon = 1$. 
We have chosen to report on this test because (similar to the 
isentropic vortex solution) its straightforward
to implement making it very accessible. Indeed, the solution
was also used as a code test by Bugner et al.~\cite{Bugner:2015gqa}
and we have confirmed our codes give nearly 
identical results~\cite{Bugner:Email:2016} (see Sec.~\ref{sec:bugner}).
Other smooth, analytic solutions, such as the relativistic vortex of
Balsara and Kim~\cite{balsara2016subluminal}, could have also been used
to test the codes's performance for smooth solutions.
Our numerical experiment, fully described in the caption of 
Table~\ref{tab:RE3D_convergence}, varies both the polynomial 
degree and number of elements. The resulting numerical errors, quoted in 
Table~\ref{tab:RE3D_convergence}, exhibit
the expected convergence rate for a fixed polynomial degree $N$
and spectral convergence with $N$. 

\begin{table*}
\setlength{\tabcolsep}{6pt}
\centering
\begin{tabular}{c | c | c | c | c | c | c | c }
$N_{\rm C}$ & $K=1$ & $K=2$ & $K=4$ & $K=8$ & $K=16$ & $K=32$ & $K=64$  \\
\hline
2 & 3.5e-01  & 3.5e-01 (0.0) & 3.1e-01 (0.2) & 1.6e-01 (1.0) & 4.4e-02 (1.8) & 1.1e-02 (2.0) & 2.8e-03 (2.0) \\
3 & 3.5e-01  & 1.6e-01 (1.2) & 1.6e-02 (3.2) & 3.8e-03 (2.1) & 7.6e-04 (2.3) & 1.2e-04 (2.6) & 1.7e-05 (2.9) \\
4 & 3.0e-01  & 5.7e-02 (2.4) & 3.5e-03 (4.0) & 1.0e-04 (5.1) & 5.3e-06 (4.3) & 3.4e-07 (3.9) & 2.1e-08 (4.1) \\
5 & 1.8e-01  & 4.3e-03 (5.4) & 1.4e-04 (5.0) & 6.9e-06 (4.3) & 3.3e-07 (4.4) & 1.3e-08 (4.7) & 4.3e-10 (4.9) \\
6 & 1.1e-01  & 1.3e-03 (6.4) & 1.2e-05 (6.7) & 9.6e-08 (7.0) & 1.6e-09 (5.9) &  ---  &  ---  \\
\end{tabular}
\caption{$h$-convergence data for the relativistic hydrodynamics 
solution~\eqref{eq:smooth_flow_re} with
$A = 0.2$, $\gamma = 5/3$, $\left(U,V,W\right) = \left(0.2,0.2,0.2\right)$,
and $k = 2 \pi$. We report $L_1$-errors computed at $T=2$ with the local 
convergence order in parentheses. The computational domain 
is periodic on $(x,y,z)\in[0,1]^3$. 
Each convergence test uses
$N_{\rm C}$ GLL collocation points per dimension per element
and varies the numerical resolution using $K\times K\times K$ elements 
with $K$ running from $2$ to $64$. 
The numerical solution is advanced forward in time with
RK3-SSP and using $\Delta t = 1.7 \times 10^{-4}$,
which results in a negligible temporal discretization error.
We show results using the Local Lax-Friedrichs numerical flux; 
other flux choices are qualitatively similar.
\label{tab:RE3D_convergence}
}
\end{table*}

\subsubsection{Cylindrical blast wave (RMHD)}
\label{sec:BlastWaveMHD}

The magnetized cylindrical blast wave test provides a test of a single
strong shock propagating in an asymmetric manner in a two-dimensional
domain. We evolve the system in 3D, assuming symmetry along the
z-axis. We consider an ideal gas equation of state $P=(\gamma -1
)\epsilon \rho$ with $\gamma=4/3$. At the initial time, we have
high-density hot material at rest within a cylinder of radius $r <
0.8$, and low-density cold material at rest at $r>1$, with a smooth
transition between the two regions.  The system is initially threaded
by a constant magnetic field $\tilde B^i = (0.1\,,0\,,0)$.  The inner
region has $\rho=0.01$, $P=1$, while the outer region has
$\rho=0.0001$, $P=0.0005$.  In the transition region $0.8<r<1$, we
perform linear interpolation in the logarithm of $\rho$ and $P$.  The
resulting evolution is a blast wave which, because of the asymmetry
introduced by the magnetic field, propagates faster along the x-axis
than the y-axis.  The computational domain covers $x=[-6,6]$,
$y=[-6,6]$, and uses $100^2\times 1$ elements with 2 GLL points along
each dimension in each element.

\begin{figure}
\centering
\includegraphics[width=1\linewidth]{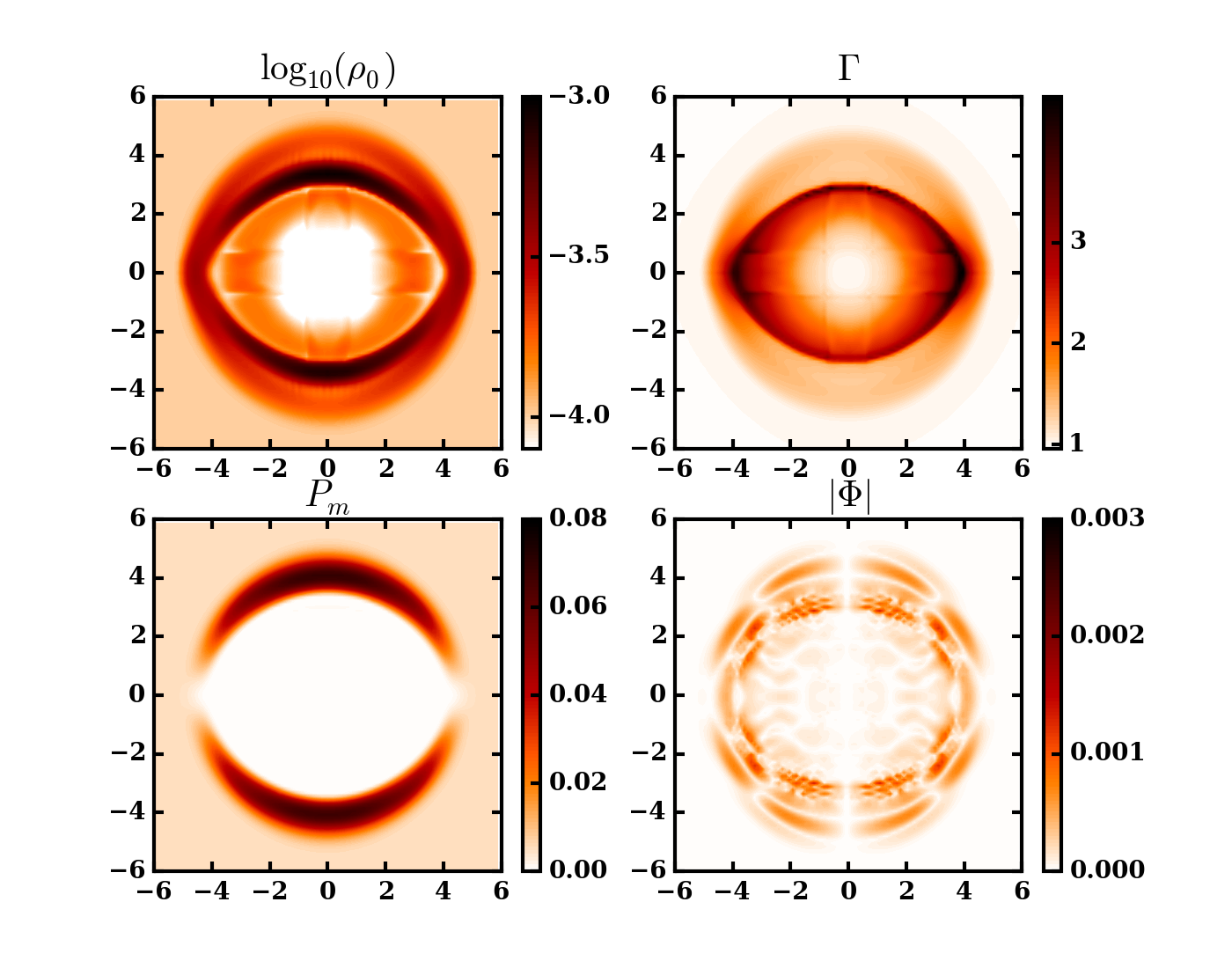}
\caption{Cylindrical blast wave test at $t=4$, using the MRS limiter
  with $\alpha = 100$. The shock wave is accurately evolved, but the
  grid is visible in some quantities behind the shock. 
  We show the fluid density $\rho_0$, Lorentz factor
  $\Gamma$, magnetic pressure $P_m=b^2/2$, and the absolute value
  of the divergence cleaning scalar $\Phi$.}
\label{fig:BlastMRS}
\end{figure}

\begin{figure}
\centering
\includegraphics[width=1\linewidth]{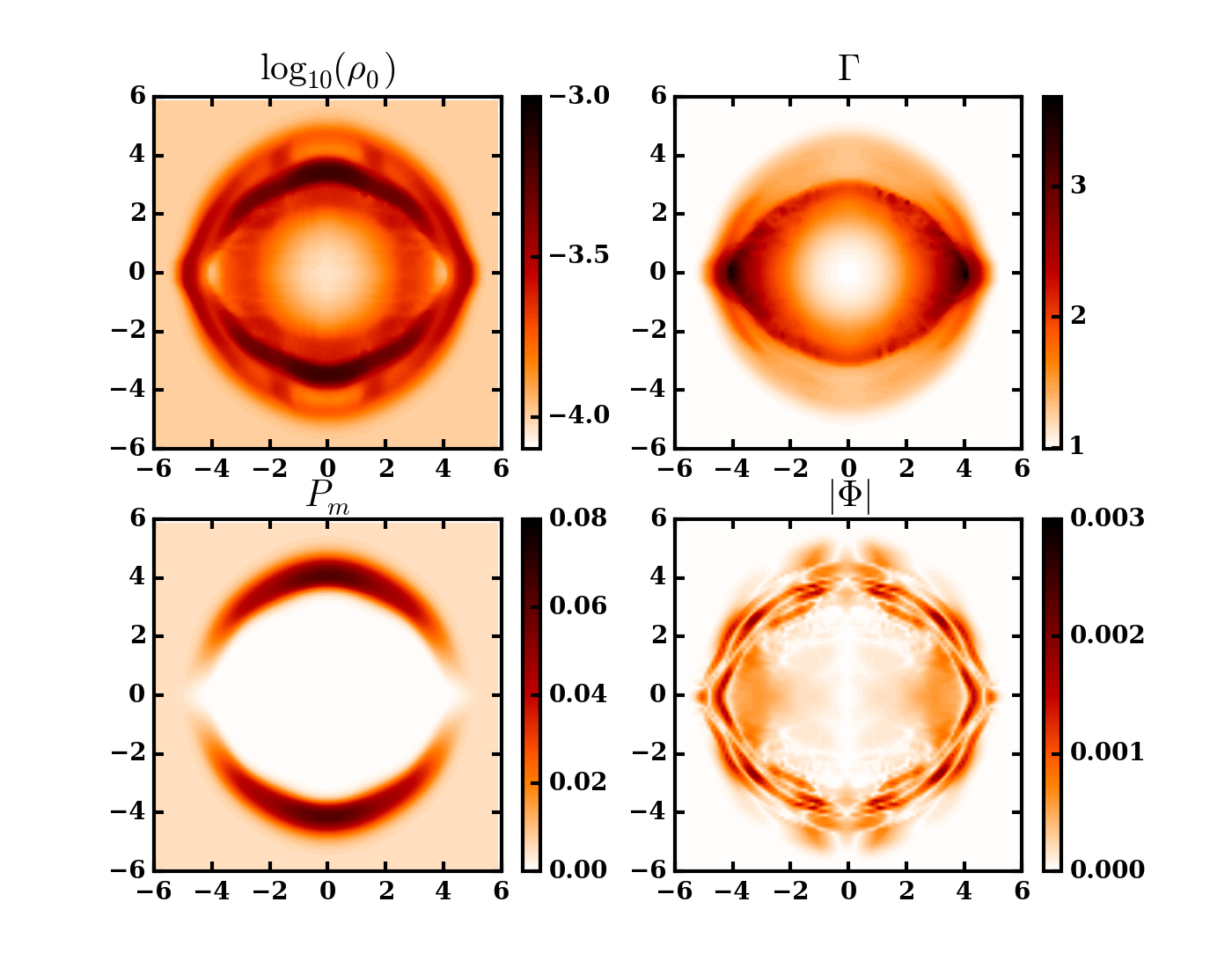}
\caption{Cylindrical blast wave test at $t=4$, using the MUSCL
  limiter. The shock wave is not as sharp as with the high-order MRS
  limiter, but does not show any structure along the coordinate axis
  of the grid.  }
\label{fig:BlastMuscl}
\end{figure}

We show results at $t=4$ in Fig.~\ref{fig:BlastMRS}, using the MRS
limiter with $\alpha=100$. These results can be compared with finite
volume results using the same number of grid points ($200 \times
200$)~\cite{DelZanna2007}. Our results are generally smoother upstream
of the blast wave and in the high density regions, but show features
associated with the grid structure behind the shock that are not
present in the finite volume simulations. If we perform the same
simulation using the MUSCL limiter (see Fig.~\ref{fig:BlastMuscl}),
the unphysical features behind the shock disappear, but the overall
quality of the solution is significantly lower. This is because the
MUSCL limiter aggressively limits the solution within each DG element,
causing the code to effectively behave like a lower-resolution
finite-volume scheme in the presence of shocks.

\subsubsection{Orszag-Tang vortex (RMHD)}
\label{sec:OTVortex}

We now consider the evolution of the
Orszag-Tang vortex benchmark test~\cite{orszag1979}, which 
features shock-shock interactions. We use the MRS limiter 
with $\alpha = 10$. To the best of our
knowledge, this test is the first application 
of the MRS limiter to the (relativistic) MHD system, and so
our results further demonstrate the robustness of this 
limiter for a new, challenging test case.

The Orszag-Tang vortex has been used extensively as a test of MHD
codes,
e.g.~\cite{1995ApJ...452..785R,1998ApJ...494..317D,2000ApJ...530..508L,beckwith2011}.
It is set on a 2-dimensional periodic domain with $0\leq x<1$ and
$0\leq y<1$ with a constant initial rest mass density $\rho_0 =
25/(26\pi)$ and specific energy density $\epsilon = 5/(8\pi)$.  We use
an ideal gas equation of state $P=(\gamma-1)\epsilon \rho$, with
$\gamma = 5/3$.  The fluid velocity is $v_x = -0.5\sin{(2\pi y)}$,
$v_y=0.5\sin{(2\pi x)}$, and the magnetic field is
$B_x=-(1/\sqrt{4\pi})\sin{(2\pi y)}$, $B_y=(1/\sqrt{4\pi})\sin{(4\pi
  x)}$. Although the Orszag-Tang vortex is a 2D problem, we formally
evolve it in 3D (but with only one DG element in the z-direction, with
periodic boundary conditions).

In Fig.~\ref{fig:OT}, we show the state of the system after $t=0.8$,
for DG elements of size $\Delta x = \Delta y = \Delta z = 0.01$ and
with $2\times 2\times 2$ collocation points each.  We find results
consistent with those of finite volume codes at similar resolution,
which is as good as one might expect for a test with multiple strong
shocks propagating and interacting.

\begin{figure}
\centering
\includegraphics[width=1\linewidth]{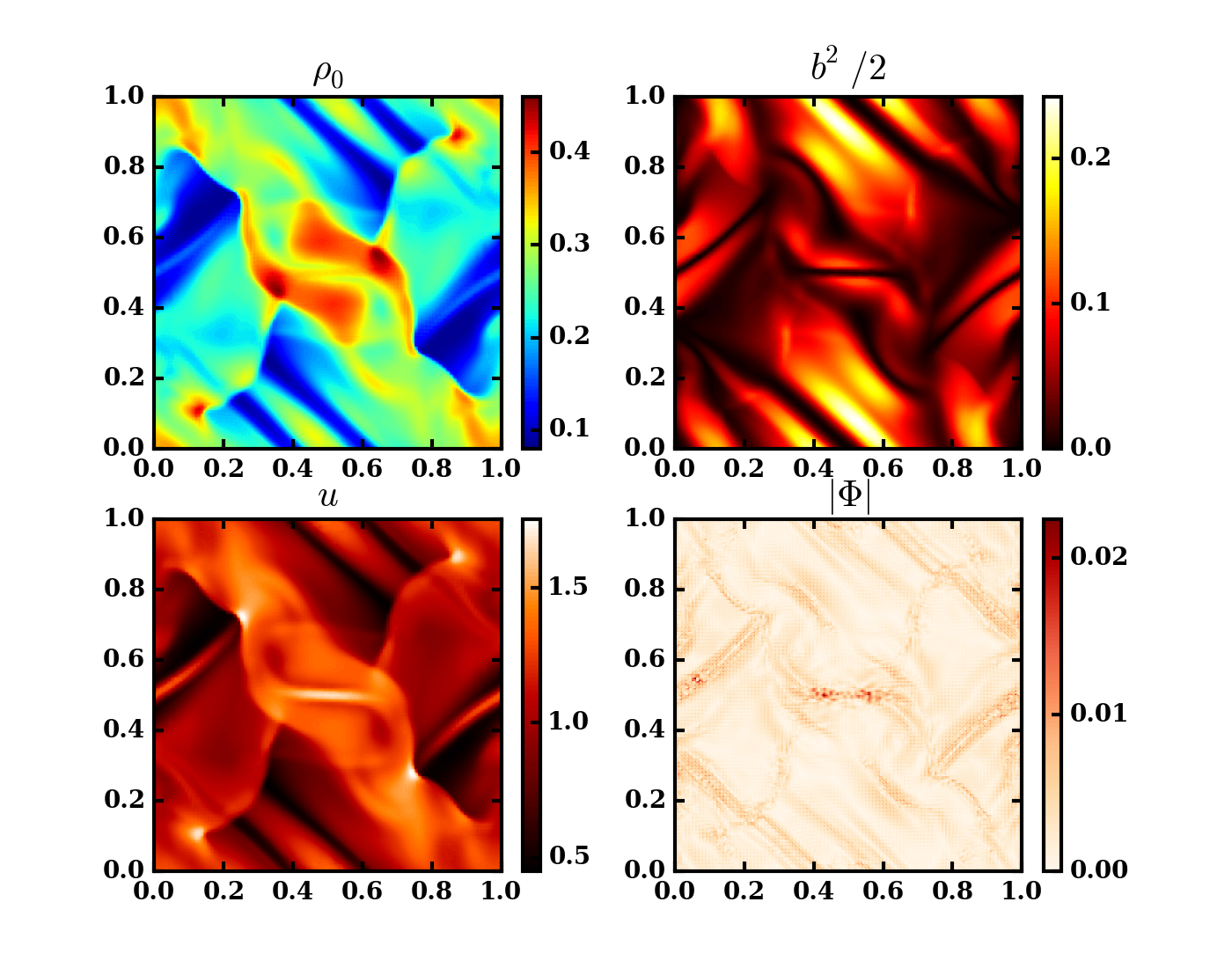}
\caption{Orszag-Tang vortex at time $t=0.8$. We show the fluid density
  (top left), the magnetic pressure (top right), the fluid internal
  energy (bottom left), and the absolute value of the
  divergence cleaning scalar $\Phi$
  (bottom right).  }
\label{fig:OT}
\end{figure} 

The Orszag-Tang test is particularly demanding for the divergence
cleaning scheme used in SpECTRE.  Numerical errors constantly drive
the divergence of the magnetic field, particularly at shocks. We see
in Fig.~\ref{fig:OT} that the divergence cleaning scalar at $t=0.8$ is
$\sim 3\%$ of the value of the magnetic field or, maybe more
meaningfully, $\sim 0.3\%$ of $|\nabla {\bf B}|$. At the same
evolution time, the maximum of $\Phi$ is $\sim 0.5\%$ of $|\nabla {\bf
  B}|$ for $\Delta x = 0.02$.

\subsubsection{Spherical Bondi accretion at $t=0$ (GRMHD)}
\label{sec:Bondi}

All of the tests considered so far were performed in flat space on a
Minkowski metric. To test the general relativistic terms in the GRMHD
equations, we consider the well-known steady-state solution for
spherical accretion onto a Schwarzschild black hole, due to
Michel~\cite{Michel:1972}. Michel's solution is the general
relativistic version of spherical Bondi accretion. Conveniently, if
one adds a radial magnetic field of the form $\tilde B^r = \tilde B^0
r^{-2}$ to that solution, it remains a steady-state solution. It is a
useful test of the GRMHD equations because many general relativistic
source terms are non-zero and have to exactly cancel each other in
order to recover a static solution. Here, we consider an accretion
flow with its sonic point at $r_S=8GM/c^2$, an ideal gas equation of
state $P = (\gamma-1)\epsilon \rho$ with $\gamma=4/3$, a radial
magnetic field $B^0=0.1$, a black hole mass $M=1$, and an accretion
rate $dM/dt=1$.  As we have not yet implemented excision of a black
hole's interior, and have not attempted to evolve singularities within
our computational domain, we do not evolve the problem. Instead, we
only verify that the time derivative of all variables converges to
zero at the initial time.

We consider two different computational domains. First, we use a
domain that does not overlap with the sonic point, $x=[4.5,5.5]r_g$,
$y=[-0.5,-0.5]r_g$, and $z = [-0.5,0.5]r_g$, with $r_g = GM/c^2$. Then
we consider a computational domain with $x=[7.5,8.5]r_g$, which
overlaps the sonic point. In both cases, we do not apply any limiting
to the solution. We find first-order convergence of the time
derivatives when using 2 basis functions per dimension in each DG
element, and third-order convergence when using 4 basis functions.

%%%%%%%%%%%%%%%%%%%%%%%%%%%%%%%%%%%%%%%%%%%%%%%%%%%%%%%%%%%%%%%%%%%%%%%%%%%%%%%
\section{Discussion}
%%%%%%%%%%%%%%%%%%%%%%%%%%%%%%%%%%%%%%%%%%%%%%%%%%%%%%%%%%%%%%%%%%%%%%%%%%%%%%%

We have described a new relativistic astrophysics code SpECTRE. 
The code differs from other codes in many important respects. In particular,
we use a discontinuous Galerkin numerical solver 
and a task-based parallelism model. To the best of our knowledge, 
this is the first DG solver parallelized using a task-based model
and the first DG treatment of the general
relativistic MHD system. As DG methods are comparatively
new to astrophysics 
(see Refs.~\cite{zanotti2015,Bugner:2015gqa,schaal2015astrophysical}
for recent work), we have provided 
a detailed description of our numerical scheme 
including those approximate Riemann solvers and high-resolution
shock capturing limiters we have implemented. 

Our main results are the performance and benchmark tests
given in \S\ref{sec:performance} 
and \S\ref{sec:tests}.  In particular,
we have shown that the code can solve
a wide variety of challenging 
astrophysics benchmark tests, including highly-relativistic shocks
with blast waves (see \S\ref{sec:RP_RE1d}), 
shock-shock interactions (see \S\ref{sec:OTVortex}),
and the steady-state GRMHD solution of Michel
(see \S\ref{sec:Bondi}).
For smooth solutions, we have shown the expected exponential
convergence with increasing grid resolution,
for example in the case of smooth relativistic flow,
\S\ref{sec:RESmooth}. In \S\ref{sec:TimeProfile}
we demonstrate how asynchronous, non-blocking communication
promotes efficient resource usage by reducing the amount of
idle core time. The scalability experiments of \S\ref{sec:strongscaling}
demonstrate the code's performance on large machines.
We observe excellent strong scaling on Blue Waters
up to the machine's full capacity of $22,380$ nodes 
using $671,400$ worker cores/threads.

Because of the generality and robustness of the DG method, it can
serve as the core kernel for the solution of multi-physics problems
that require an ensemble of interacting physical descriptions.  Work
toward this is underway.  To evolve dynamical spacetimes we are
implementing the Einstein equations using a generalized harmonic
formulation~\cite{Pretorius2005c,Gundlach2005} written with
first-order time and first-order space
derivatives~\cite{Lindblom:2007}.  We consider the fully first-order
generalized harmonic system because it has been successfully evolved
with pseudospectral methods for a variety of astrophysical
configurations~\cite{Haas:2016,SpECwebsite}.  To tackle challenging
multi-scale problems, we are adding local time stepping techniques and
mesh refinement strategies to either split the elements into smaller
elements ($h$-adaptivity) or increase the number of basis functions in
an element ($p$-adaptivity).  The locality of the DG scheme
facilitates adaptive resolution refinement and local timestepping, and
good results have been shown in other contexts.  And while current
large codes often struggle to achieve good performance with local
time-stepping~\cite{dubey14}, task-based parallelism and asynchronous
communication are promising tools to overcome this problem.

Looking forward, we believe that the benefits of
task-based parallelism will become 
increasingly important in these more complicated contexts. 
More generally, accurate numerical methods and efficient usage of massively 
parallel supercomputers
will be essential for the high-fidelity simulations needed to realize the
promise of current and future experiments.

%%%%%%%%%%%%%%%%%%%%%%%%%%%%%%%%%%%%%%%%%%%%%%%%%%%%%%%%%%%%%%%%%%%%%%%%%%%%%%%
% Acknowledgments
%%%%%%%%%%%%%%%%%%%%%%%%%%%%%%%%%%%%%%%%%%%%%%%%%%%%%%%%%%%%%%%%%%%%%%%%%%%%%%%
\section*{Acknowledgments}

\noindent We acknowledge helpful discussions with
Sebastiano Bernuzzi, 
Bernd Br\"ugmann, 
Marcus Bugner, 
Mani Chandra, 
Roland Haas, 
Daniel Hemberger, 
Jan Hesthaven, 
Cameron Hummels, 
Fraser Hutchison,
Jonathan Lifflander, 
Geoffrey Lovelace, 
Phil Miller, 
Harald Pfeiffer, 
David Radice, 
and Christian Reisswig. 
We thank the anonymous referee for detailed
comments and suggestions which helped to improve the manuscript.
Support for F. Foucart was provided by NASA through Einstein
Postdoctoral Fellowship grant numbered PF4-150122 awarded by the
Chandra X-ray Center, which is operated by the Smithsonian
Astrophysical Observatory for NASA under contract NAS8-03060.  Authors
at Cornell were partially supported by NSF under award nos.\ TCAN
AST-1333129 and PHY-1306125, and by the Sherman Fairchild Foundation.
Authors at Caltech were partially supported by NSF under award
nos.\ TCAN AST-1333520, CAREER PHY-1151197, and PHY-1404569, and by
the Sherman Fairchild Foundation.  ES acknowledges support from NSF
award OCI-0905046.  ES and JM were supported by an NSERC Discovery
grant.  Research at the Perimeter Institute is supported by the
Government of Canada through Industry Canada and by the Province of
Ontario through the Ministry of Research \& Innovation.  Computations
were performed on NSF/NCSA Blue Waters under allocation PRAC
ACI-1440083, on the NSF XSEDE network under allocations TG-PHY100033
and TG-PHY990007, and on the Caltech compute cluster Zwicky (NSF
MRI-R$^2$ award no.\ PHY-0960291).

\appendix
\makeatletter
\renewcommand\thesubsection{\@Alph\c@section.\@arabic\c@subsection}
\makeatother

%%%%%%%%%%%%%%%%%%%%%%%%%%%%%%%%%%%%%%%%%%%%%%%%%%%%%%%%%%%%%%%%%%%%%%%%%%%%%%%
\section{Performance of fluxes and limiters for the Sod problem}
\label{app:sod}
%%%%%%%%%%%%%%%%%%%%%%%%%%%%%%%%%%%%%%%%%%%%%%%%%%%%%%%%%%%%%%%%%%%%%%%%%%%%%%%
\setcounter{table}{0}
\setcounter{figure}{0}

\begin{table*}
\setlength{\tabcolsep}{6pt}
\centering
\begin{tabular}{c | c | c | c | c | c | c}
Scheme & $n_x=20$ & $n_x=40$ & $n_x=80$ & $n_x=160$ & $n_x=320$ & $n_x = 640$ \\
\hline
HLL-MinMod-2 & 1.7e-01  & 9.5e-02 (0.9) & 4.9e-02 (1.0) & 2.6e-02 (0.9) & 1.4e-02 (0.9) & 7.7e-03 (0.9) \\
Roe-MinMod-2 & 1.7e-01  & 9.3e-02 (0.9) & 4.8e-02 (1.0) & 2.6e-02 (0.9) & 1.4e-02 (0.9) & 7.5e-03 (0.9) \\
LLF-MinMod-2 & 1.9e-01  & 1.1e-01 (0.8) & 5.4e-02 (1.0) & 2.9e-02 (0.9) & 1.5e-02 (0.9) & 8.5e-03 (0.9) \\
Mar-MinMod-2 & 1.9e-01  & 1.0e-01 (0.9) & 5.3e-02 (1.0) & 2.8e-02 (0.9) & 1.5e-02 (0.9) & 8.0e-03 (0.9) \\
\hline
Roe-MinModN-3 & 1.2e-01  & 6.4e-02 (0.9) & 3.3e-02 (1.0) & 1.7e-02 (1.0) & 8.3e-03 (1.0) & 4.4e-03 (0.9) \\
Roe-MRS0.5-3 & 1.3e-01  & 6.4e-02 (1.0) & 3.3e-02 (1.0) & 1.7e-02 (1.0) & 8.6e-03 (0.9) & 4.6e-03 (0.9) \\
Roe-MRS5-3 & 1.1e-01  & 6.0e-02 (0.9) & 3.1e-02 (0.9) & 1.6e-02 (1.0) & 8.2e-03 (1.0) & 4.3e-03 (0.9) \\
Roe-MRS50-3 & 6.5e-02  & 3.9e-02 (0.7) & 2.8e-02 (0.5) & 1.5e-02 (0.9) & 8.0e-03 (0.9) & 4.2e-03 (0.9) \\
Roe-MinModN-4 & 1.3e-01  & 6.7e-02 (1.0) & 3.5e-02 (0.9) & 1.8e-02 (1.0) & 8.8e-03 (1.0) & 4.6e-03 (0.9) \\
Roe-MRS0.5-4 & 1.2e-01  & 6.7e-02 (0.9) & 3.6e-02 (0.9) & 1.9e-02 (0.9) & 1.1e-02 (0.8) & 5.2e-03 (1.0) \\
Roe-MRS5-4 & 1.2e-01  & 6.3e-02 (1.0) & 3.2e-02 (1.0) & 1.8e-02 (0.9) & 9.0e-03 (1.0) & 4.9e-03 (0.9) \\
Roe-MRS50-4 & 5.7e-02  & 4.6e-02 (0.3) & 2.8e-02 (0.7) & 1.5e-02 (0.9) & 8.8e-03 (0.8) & 4.6e-03 (0.9) \\
\end{tabular}
\caption{$h$-convergence data for the NE-Sod problem. 
We report $L_1$-errors computed at $T=0.25$ with the local 
convergence order in parentheses. The computational domain 
$x\in[-0.5\,,0.5]$ is divided into $n_x$ uniform elements.  
The numerical solution is advanced forward in time with RK3-SSP. 
\label{tab:NE_sod}
}
\end{table*}

In \S\ref{sec:RP_NE1d} we considered
the standard Sod shock tube test
whose initial data is given in Table~\ref{tab:Riemann_Problems}.
In this appendix we summarize the approximately 80
$h$-convergence simulations we performed, from which we 
draw a few general insights into the  
relative strengths 
and weaknesses of different limiters and numerical fluxes.
Table~\ref{tab:NE_sod} reports the numerical errors
compared to the exact solution. In all cases, our numerical scheme
exhibits a convergence rate of $\approx 1$, as
expected for discontinuous solutions of this type.

Our first numerical test explores 
the impact of numerical flux choice when using low
order elements (degree $N=1$ basis functions) and a minmod limiter.
Numerical errors (see Table~\ref{tab:NE_sod}) 
are computed for four different numerical 
fluxes and show all four schemes perform remarkably similar. 
The numerical solutions computed with these 
schemes also appear visually identical; one representative
case is show in  the upper-left panel of Fig.~\ref{fig:Sod1D_CompareLimiter}.
These findings are consistent with those of
Qiu and Shu~\cite{qiu2005comparison}.

The next numerical test considers 
the behavior of the higher-order limiters
when using higher-order elements.  
As demonstrated in Table.~\ref{tab:NE_sod}, these limiters 
continue to offer accuracy comparable to their lower order
counterparts. This is anticipated, since, 
for a discontinuous solution of this type, 
linear basis elements are expected to be optimal. 
Higher elements may provide marginal accuracy benefits at best.
At worst, higher elements may lead to 
large overshoots. Varying the numerical flux,
limiter, and polynomial order, we have explored $\approx 80$
different convergence tests in total, and 
find most (reasonable) combinations 
give consistently similar numerical errors
(see Table.~\ref{tab:NE_sod}).

Visually inspecting the numerical solution's ``quality"
provides another important measure of success. Looking 
at a $T=0.25$ snapshot now reveals significant differences between
schemes, which are not as easily
captured in a numerical error computation. Our most interesting 
observations are summarized in Fig.~\ref{fig:Sod1D_CompareLimiter}
and its caption. When combined with a HRSC limiter,
we consistently find that the Roe flux outperforms
all three of its competitors while the Local Lax Friedrichs flux
typically performs the worst. When used with a Roe
flux, we find that the HRSC limiters
described in \S\ref{sec:HRSCLimiters}
perform very well on this test
case. Note that the authors of ref.~\cite{schaal2015astrophysical}
report 
good shock capturing behavior when using a LLF flux
combined with a characteristic variable limiting procedure,
which we have not explored here.

\begin{figure}
\includegraphics[width=0.99\linewidth]{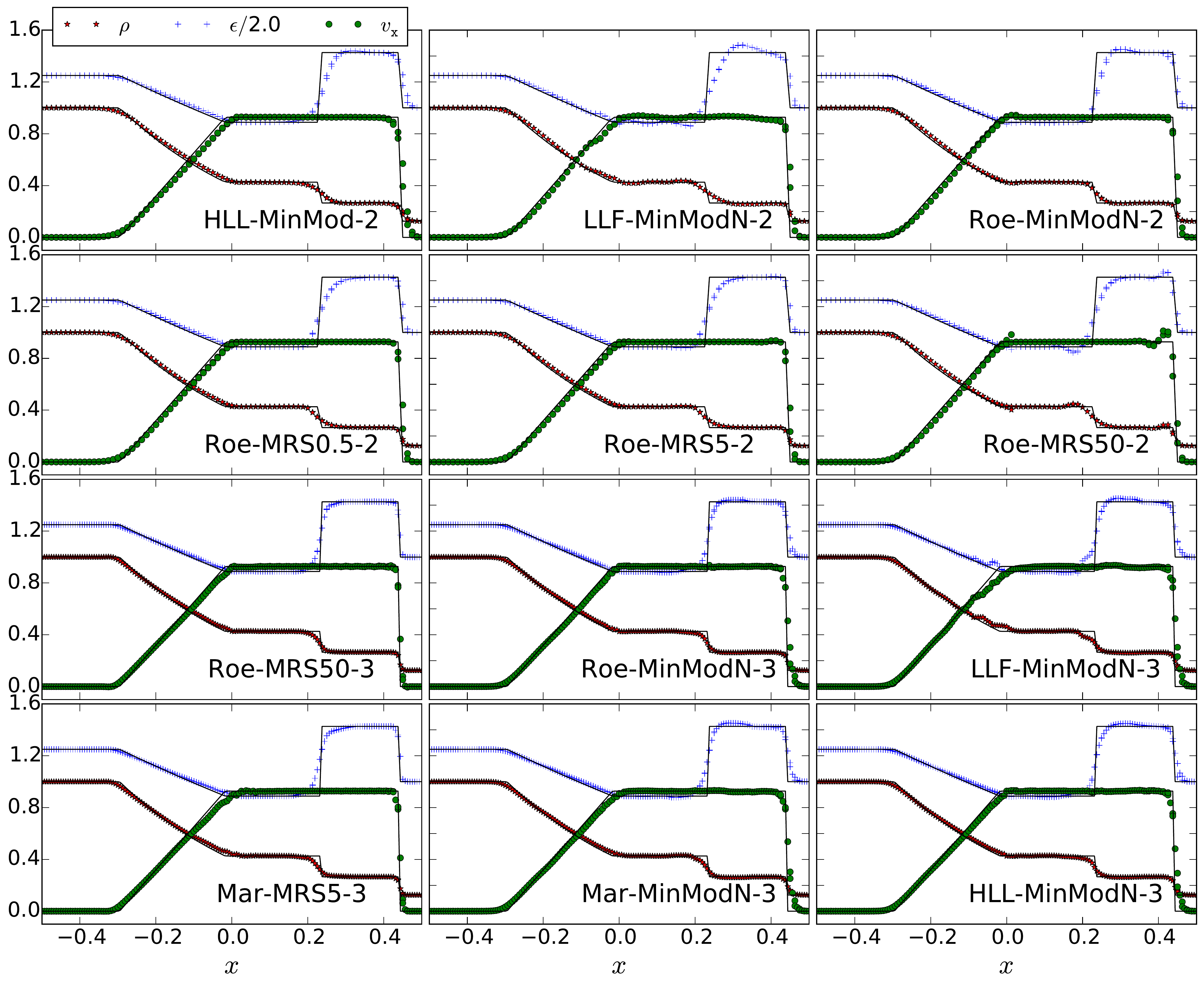}
\caption{Comparison of the Sod Shock tube problem computed with 
different limiters and numerical fluxes. The setup is described 
in the caption of Table.~\ref{tab:NE_sod}.
We show results with a moderate grid resolution
using $80$ uniform elements. 
{\bf Top row}: The first panel shows a typical minmod solution, 
which is fairly consistent across numerical flux choice. The 
next two panels show the performance of a high-$N$ minmod
limiter. Despite using only first order elements, noticeable 
oscillations appear in both simulations, with the LLF
flux containing significantly more. 
Qualitatively, the HLL and Marquina fluxes (not shown) 
appear more similar to the LLF case. 
{\bf Second row}: These three panels depict a sequence 
of solutions computed with the parameterized MRS limiter using 
$\alpha = 0.5\,,5\,,50$. We observe the limiter to be
effective over a large range of values, although as 
$\alpha$ becomes large the limiter turns off
and oscillations return.
{\bf Third row}: In the first two panels we see that both 
the high-N minmod and MRS limiters may perform {\em better}
than their lower-order counterparts.
For the LLF flux (third panel) this is not the case, unfortunately.
{\bf Bottom row}: The first two panels use a HRSC limiter with
the Marquina flux and second order elements. The solution quality is
seen to be not quite as good as the comparable simulations using 
a Roe flux, but still significantly
better than the LLF and fairly good overall. The bottom right panel,
together with its three neighboring panels, shows all four numerical fluxes
using second order elements and the high-N minmod limiter.}
\label{fig:Sod1D_CompareLimiter}
\end{figure}

\FloatBarrier     %keep next set of tables in the right place
%%%%%%%%%%%%%%%%%%%%%%%%%%%%%%%%%%%%%%%%%%%%%%%%%%%%%%%%%%%%%%%%%%%%%%%%%%%%%%%
\section{Numerical truncation error and convergence tables}
\label{app:tables}
%%%%%%%%%%%%%%%%%%%%%%%%%%%%%%%%%%%%%%%%%%%%%%%%%%%%%%%%%%%%%%%%%%%%%%%%%%%%%%%
\setcounter{table}{0}

A table of numerical data provides an effective means for comparing codes. 
One particularly important use is to verify that an algorithm has been 
correctly implemented. Since there are very few relativistic hydrodynamics 
codes using a discontinuous Galerkin solver, 
there are correspondingly few tables to compare with. We were also 
hard pressed to find any tables with which to compare the isentropic vortex 
(non-relativistic Euler) solution since, in many other fields, the 
DG scheme is formulated on tetrahedra and with a full mass matrix computed
using Gaussian quadratures. As discussed in \S\ref{sec:algorithm}, our scheme
uses a GLL quadrature rule that results in a diagonal mass
matrix~\cite{gassner2011comparison,teukolsky2015short}. In this 
appendix we provide a sample of tables of errors and $h$-convergence data
for tests considered in this paper.

\begin{table*}
\setlength{\tabcolsep}{6pt}
\centering
\begin{tabular}{c | c | c | c | c | c | c}
$N_{\rm C}$ & $n_x=10$ & $n_x=20$ & $n_x=40$ & $n_x=80$ & $n_x=160$ & $n_x=320$  \\
\hline
2 & 2.55e-02  & 6.60e-03 (1.9) & 1.66e-03 (2.0) & 4.16e-04 (2.0) & 1.04e-04 (2.0) & 2.60e-05 (2.0) \\
3 & 1.07e-03  & 2.02e-04 (2.4) & 3.07e-05 (2.7) & 4.10e-06 (2.9) & 5.21e-07 (3.0) & 6.54e-08 (3.0) \\
4 & 1.24e-05  & 7.88e-07 (4.0) & 4.57e-08 (4.1) & 2.81e-09 (4.0) & 1.75e-10 (4.0) & 1.19e-11 (3.9) \\
5 & 1.17e-06  & 5.27e-08 (4.5) & 1.94e-09 (4.8) & 6.38e-11 (4.9) & 3.67e-12 (4.1) & 2.81e-12 (0.4)
\end{tabular}
\caption{$h$-convergence data for the one-dimensional version of the
  smooth flow solution~\eqref{eq:smooth_flow_re} with $A = 0.2$,
  $\gamma = 5/3$, $v_x = 0.2$, and $k = 2 \pi$.  We report
  $L_1$-errors computed at $T=2$ with the local convergence order in
  parentheses. The computational domain is periodic on $x\in[0,1]$ and
  divided into $n_x$ uniform elements.  Each convergence test uses
  $N_{\rm C}$ GLL collocation points per dimension per element.  The
  numerical solution is advanced forward in time with RK3-SSP using
  $\Delta t = 10^{-4}$, which results in a negligible temporal
  discretization error.  We show results using the Local
  Lax-Friedrichs numerical flux; other flux choices are qualitatively
  similar.  Our results exactly match those produced by the code of
  Bugner et al (see Sec.~\ref{sec:bugner}) up 
  to $\approx 5\times10^{-12}$. We believe
  that discrepancies observed in smaller values of the numerical error
  are likely due to a difference in timesteppers and/or the effect of
  roundoff errors.
\label{tab:Bugner_comparison}
}
\end{table*}

\begin{table*}
\setlength{\tabcolsep}{6pt}
\centering
\begin{tabular}{l l | c | c | c | c | c}
Test & Scheme & $n_x=40$ & $n_x=80$ & $n_x=160$ & $n_x=320$ & $n_x=640$ \\
\hline
ST1 & HLL-MinMod-2{}{} & 7.1e-02 (0.7) & 4.8e-02 (0.6) & 2.6e-02 (0.9) & 1.4e-02 (0.9) & 7.5e-03 (0.9) \\
    & HLL-MinModN-2{} & 5.8e-02 (0.8) & 3.5e-02 (0.7) & 1.9e-02 (0.9) & 1.0e-02 (0.9) & 6.2e-03 (0.7) \\
    & Mar-MRS0.5-2{}{} & 8.2e-02 (0.6) & 5.1e-02 (0.7) & 2.5e-02 (1.0) & 1.1e-02 (1.2) & 5.6e-03 (1.0) \\
    & Roe-MRS5-3 & 6.8e-02 (0.5) & 4.3e-02 (0.6) & 1.9e-02 (1.2) & 6.5e-03 (1.5) & 3.3e-03 (1.0) \\[2pt]
\hline
 &  & $n_x=80$ & $n_x=160$ & $n_x=320$ & $n_x=640$ & $n_x=1280$ \\
\hline
ST2 & LLF-MinMod-2 & 5.1e-02 (0.7) & 3.1e-02 (0.7) & 1.9e-02 (0.7) & 1.1e-02 (0.7) & 6.9e-03 (0.7) \\
    & LLF-MinModN-2 & 9.8e-02 (0.7) & 5.9e-02 (0.7) & 3.3e-02 (0.8) & 2.1e-02 (0.6) & 1.4e-02 (0.6) \\
    & Roe-MRS50-3 & 2.4e-02 (0.5) & 1.5e-02 (0.6) & 1.0e-02 (0.6) & 7.0e-03 (0.5) & 6.2e-03 (0.2) \\
    & Mar-MRS5-2 & 4.3e-02 (0.7) & 2.5e-02 (0.8) & 1.6e-02 (0.6) & 1.1e-02 (0.6) & 7.3e-03 (0.6)
\end{tabular}
\caption{Sample of $h$-convergence data for the relativistic Euler
  Riemann problems shock tube 1 (ST1) and shock tube 2 (ST2). Cases
  not shown are qualitatively similar.  The ST2 test is the more
  challenging of the two, with a narrow blast wave feature and Lorentz
  factors of about $6$. Because of the large size of the exact
  solutions, we report {\em relative} $L_1$-errors computed at $T=0.4$
  with the local convergence order in parentheses. The normalization
  factor is simply the exact solution's norm at the final time as
  computed on the numerical grid, $\| u \|_{L_1} \approx 21$ for ST1
  and $\| u \|_{L_1} \approx 1,151$ for ST2. The computational domain
  $x\in[0,1]$ is divided into $n_x$ uniform elements.  The numerical
  solution is advanced forward in time with RK3-SSP.
To check our implementation of the various limiters and numerical fluxes,
we have performed a total of $\approx 50$ convergence studies for each 
test and find the numerical error for comparable values of $n_x$ to be at most
a factor of a few different from the cases shown here.
\label{tab:RE_sod}
}
\end{table*}

\begin{table*}
\setlength{\tabcolsep}{6pt}
\centering
\begin{tabular}{c | c | c | c | c | c | c}
%% LLF %%
$N_{\rm C}$ & $K=8$ & $K=16$ & $K=32$ & $K=64$ & $K=128$ & $K=256$ \\
\hline
2 & 1.51e+01  & 5.59e+00 (1.4) & 1.73e+00 (1.7) & 4.62e-01 (1.9) & 1.17e-01 (2.0) & 2.93e-02 (2.0) \\
3 & 2.92e+00  & 6.29e-01 (2.2) & 1.02e-01 (2.6) & 1.78e-02 (2.5) & 3.21e-03 (2.5) & 5.83e-04 (2.5) \\
4 & 1.00e+00  & 9.73e-02 (3.4) & 6.70e-03 (3.9) & 3.44e-04 (4.3) & 2.15e-05 (4.0) & 1.41e-06 (3.9) \\
5 & 2.37e-01  & 7.98e-03 (4.9) & 3.13e-04 (4.7) & 1.44e-05 (4.4) & 6.94e-07 (4.4) & 3.66e-08 (4.2) \\
6 & 4.95e-02  & 1.37e-03 (5.2) & 2.21e-05 (5.9) & 3.03e-07 (6.2) & 5.38e-09 (5.8) & 2.45e-10 (4.5) \\
7 & 1.15e-02  & 9.88e-05 (6.9) & 9.32e-07 (6.7) & 1.10e-08 (6.4) & 2.68e-10 (5.4) &  ---  \\
8 & 2.21e-03  & 1.41e-05 (7.3) & 7.46e-08 (7.6) & 3.57e-10 (7.7) &  ---  &  ---
\end{tabular}
\caption{$h$-convergence data for the isentropic vortex
  solution~\eqref{eq:isen_vortex} with $\beta = 5$, $X_0 = Y_0 = 4$,
  $\left(U,V,W\right) = \left(1, 1, 0 \right)$, and $\gamma = 1.4$. We
  report $L_1$-errors computed at $T=2$ with the local convergence
  order in parentheses.  The physical domain is taken to be the slab
  $(x,y) \in [0,10]$ and $z \in [0,1]$ with analytic outer boundary
  conditions.  Each convergence test uses $N_{\rm C}$ GLL collocation
  points per dimension per element and varies the numerical resolution
  using $K\times K\times 1$ elements with $K$ running from $8$ to
  $256$.  The numerical solution is advanced forward in time with
  RK3-SSP using $\Delta t = 1 \times 10^{-4}$, which results in a
  temporal discretization error of $\approx 2 \times 10^{-10}$.  We
  show results using the LLF numerical flux; other flux choices lead
  to qualitatively similar results.
\label{tab:IsenVortexConvergence}
}
\end{table*}

\newpage

%%%%%%%%%%%%%%%%%%%%%%%%%%%%%%%%%%%%%%%%%%%%%%%%%%%%%%%%%%%%%%%%%%%%%%%%%%%%%%%
\section*{References}
%%%%%%%%%%%%%%%%%%%%%%%%%%%%%%%%%%%%%%%%%%%%%%%%%%%%%%%%%%%%%%%%%%%%%%%%%%%%%%%
\bibliography{References/References,References/cs,References/dg}

\end{document}